\documentstyle[12pt,aaspp4]{article}
\tightenlines

\slugcomment{}

\lefthead{Maiolino et al.}
\righthead{Seyfert Activity and Star Formation in Circinus}

\begin{document}

\title{Seyfert Activity and Nuclear Star Formation in the Circinus Galaxy}

\author{R.~Maiolino\altaffilmark{1,2}, A.~Krabbe\altaffilmark{1,3},
 N.~Thatte\altaffilmark{1}, R.~Genzel\altaffilmark{1}}

\altaffiltext{1}{Max-Plank-Institut f\"{u}r Extraterrestrische Physik,
Postfach 1603, D-85740 Garching, Germany}
\altaffiltext{2}{Osservatorio Astrofisico di Arcetri, L.go E.Fermi 5,
 I-50125 Firenze, Italy}
\altaffiltext{3}{DLR-Institute for Space Sensor Technology, Rudower Chaussee 5,
  12489 Berlin, Germany}

\begin{abstract}

We present high angular resolution ($0.15''-0.5''$)
near infrared images and spectroscopy of the Circinus galaxy,
the closest Seyfert 2 galaxy known.
The data reveal a
non--stellar nuclear source at 2.2$\mu$m
whose size is smaller than 1.5 pc in radius.
The coronal line region and the hot molecular gas emission extend for
20--50 pc in the ionization cone.
The data do not show evidence for a point--like concentration of dark mass;
we set an upper limit of $4\times 10^6 M_{\odot}$ to the mass of
a putative black hole.

We find evidence for
a young nuclear stellar population, with typical ages between 
$4\times 10^7$ and $1.5\times 10^8$ yrs. The luminosity of the
starburst inside a few hundred pc is comparable to the
intrinsic luminosity of the Seyfert nucleus, and the two of them together
account for most of the observed bolometric luminosity of the galaxy.
Within the central 12 pc the starburst has an age of about
$7\times 10^7$ yrs and radiates $\sim$ 2\%
of the luminosity of the active nucleus.
We discuss the implications of these results for
models that have been proposed for the starburst--AGN connection.

\end{abstract}


\section{Introduction} \label{intro}

The connection between starburst and Seyfert activity
is one of the most hotly debated issues in the context of AGNs.
A variety of data have provided growing evidence that starburst and
Seyfert phenomenon often coexist (Cid Fernandes \& Terlevich 1995,
Delgado and Perez 1993, Genzel et al. 1995,
Heckman et al. 1989, 1995, Maiolino et al. 1995,
Moorwood et al. 1996c, Neff et al. 1994,
Oliva et al. 1995, Rodriguez-Espinosa et al. 1987,
Storchi-Bergmann et al. 1996, Cid Fernandes et al. 1997). However, most of
these
studies have detected star forming activity, and/or a young stellar population
in Seyfert galaxies (especially type 2) on the kpc scale, which might not be
directly relevant to the active nucleus. Indeed, Maiolino et al. (1997)
speculate that star formation in the host galaxy of Seyfert 2 nuclei
is probably related only indirectly to the active nucleus: non--axisymmetric
morphologies (bars/distortions/interactions)
are likely to be the prime cause of the relationship, as they
both enhance star formation in the galaxy and drive gas into the nuclear
region to obscure the active nucleus.
Terlevich and collaborators (Terlevich \& Melnick 1985,
Terlevich et al. 1992, 1995,
Terlevich 1994) have proposed a much closer link between nuclear
star formation and Seyfert activity. They claim that the whole phenomenology
observed in AGN can be accounted for by assuming that these nuclei
are powered by a massive nuclear starburst evolving in the metal rich
environment of early type galaxies: the ionizing radiation emitted by an extreme
population of Wolf--Rayet stars, along with fast shocks generated by SNe,
would produce the spectrum observed in Sy galaxies.
So far, very few observational tests of this theory have been made,
due to difficulties in probing star formation and
stellar populations in the innermost region of
AGNs. However, recently Heckman et al. (1997) have
discovered a powerful, young (in the Wolf-Rayet phase)
nuclear starburst in the Sy2 galaxy Mkn477; such a nuclear
starburst might dominate the bolometric luminosity of the central region,
therefore partly supporting Terlevich's scenario, at least in this object.
Norman \& Scoville (1988), following Bailey(1980) and David (1987a,b),
have proposed a
different link between the nuclear stellar population
 and AGN. According to their model, a
young nuclear ($\le$ 10 pc) star cluster would return mass into the
interstellar medium, by means of post-main-sequence stars,
that would feed a nuclear black hole. An analogous model has been proposed
more recently by Murphy et al. (1991), who also include effects of tidial
disruption of stars close to the black hole.
Jenkins \& Binney (1994) have proposed a similar mechanism for the
fueling of the Galactic center.  Again,
these models have not been tested, so far, due to the shortage
of observations that can probe the nuclear stellar population.

The Circinus galaxy is a nearby (4 Mpc), edge--on (incl.$\sim$65$^\circ$),
Sb--d system that is seen through a low interstellar extinction
 window near the
Galactic plane ($A_V$ = 1.5 mag, Freeman et al. 1977).
The nuclear 
optical line ratios (Oliva et al. 1994) are typical of a Seyfert 2 galaxy.
The Seyfert 2 classification is supported also by the detection of
intense
coronal lines (Oliva et al. 1994, Moorwood et al. 1996a), the discovery of
an intense X--ray iron 6.7 keV line (Matt et al. 1996), rapid variation of
the powerful H$_2$O maser emission (Greenhill et al. 1997) and a prominent
ionization cone in the [OIII]$\lambda$5007 maps (Marconi et al. 1995), with
filamentary supersonic outflows (Veilleux \& Bland-Hawthorn 1997).
The large equivalent width of the  Fe 6.7 keV line detected by ASCA
(EW $\sim$ 2 keV) indicates that the nucleus is heavily
absorbed along the line of sight:
$N_H> 10^{25} cm^{-2}$, i.e it is Compton thick. The Circinus galaxy
is also characterized by enhanced star forming activity: H$\alpha$ and
[SII] narrow band images
(Marconi et al. 1995) have
revealed the presence of a star forming ring at a radius
of $\sim$10$''$=200pc.
Given its proximity and the coexistence of both star formation and an active
nucleus, the Circinus galaxy appears
to be one of the most suitable objects to tackle the starburst--AGN
connection issue.

In this paper we present high angular resolution near--IR images and
integral field spectroscopic data that probe the physics of the
nuclear region of the Circinus galaxy, on scales from a few parsecs to
$\sim$ 100 pc. Using these
data, we investigate the properties of the active nucleus and its
interaction with the circumnuclear environment, with emphasis on
the connection between Seyfert and stellar
activity.

\section{Observations} \label{obs}

\subsection{3D integral field spectroscopy}

The nucleus of the Circinus galaxy was observed in March and April 1996
with 3D, the MPE
near--IR imaging spectrometer (Weitzel et al. 1996), assisted by
ROGUE, a first order adaptive optics system (Thatte
et al. 1995), at the 2.2m MPE--ESO telescope.
3D slices the focal plane into 16 slits
and disperses their light in wavelength; the spectrum of the slices is then
imaged onto
a NICMOS3 detector (256$\times$256 pixels). The resulting ``cube'' provides
simultaneous spectra of an area
$4.8''\times 4.8''$, as projected on the sky, divided into 16$\times$16 pixels
(i.e. $0.3''$/pix).
We observed Circinus in two
different spectral configurations: in March (two nights)
 we observed the whole K band at a
spectral resolution of R=1000, while in April (four nights)
we observed the spectral region
around the CO stellar bands (2.2--2.4$\mu m$) at a spectral resolution
of R=2000.
The FWHM of the
point spread function (PSF) during the observations, after the tip tilt
correction, was about $0.5''-0.6''$ in the K band. One of the nights in April
had an optical seeing of $0.4''$, while the seeing in the K band, after tip tilt
correction, was about $0.35''$, though we could not determine
it accurately because
of heavy undersampling of the PSF. We observed a nearby O
star to correct the atmospheric transmission features in the R=1000 data,
as O stars are almost featureless in the K band, while we observed an A
star to correct the R=2000 data, as A stars are featureless in the 
corresponding spectral region.
Data reduction was performed as described in Weitzel et al. (1996) and
Thatte et al. (1997).

\subsection{Speckle observations}

We also observed the Circinus galaxy by means of
SHARP, the MPE speckle near-IR imaging camera at the NTT ESO telescope,
in April 1996. This camera uses a NICMOS3 array with a pixel scale
of 0.05$''$/pix (providing a field of view of $12.8''\times12.8''$).
We observed the central region of Circinus in H and K bands at
a frame rate of 1 Hz. The observations of Circinus
were interleaved with observations of a reference star (about the
same magnitude as the Circinus nucleus) to monitor the PSF. The individual
images were background subtracted, flat fielded, shifted to center the
brightest speckle and then co-added.
Details on the data reduction are given in Eckart et al.
(1993, 1995).
The K band images have a PSF of 0.2$''$ FWHM;
however, if the diffuse, seeing
limited PSF is subtracted, the central spike is close to the diffraction limit
($\sim 0.15''$ FWHM). In H, the PSF is slightly worse: 0.27$''$ FWHM.

\section{Results} \label{res}

This section is aimed at showing general results extracted from the data,
to provide an overview of the features observed in the central region
of the Circinus galaxy. A more detailed analysis is given in
Sect.4 \& 5.

\subsection{Near--IR images}

Fig.1a shows the K--band SHARP image of the central region of
Circinus,
smoothed with a 3 pixel
(0.15$''$) FWHM gaussian to improve the S/N on the low surface brightness
features; while Fig.1b shows the central 3$''$ at full resolution.
The K band image shows a two armed spiral--like morphology. The western arm is
also apparent in the 7000 \AA $~$image (Marconi et al. 1995),
but the southern arm is less evident
in the optical image, which might be partly due to obscuration towards the
South--East (see below) and partly to lack of angular resolution in the
optical images. In the central region (R$\le$80 pc)
the surface brightness distribution
is smoother and rises slowly toward the
K peak, with a shallow power law (surface brightness
$\propto R^{-\alpha}$ with $\alpha \sim$
0.5-0.7).
This central
region (R$\le$ 80 pc) bay be considered as the inner part of a bulge
(a ``nuclear bulge'', in analogy with our Galaxy, Mezger et al. 1996),
though in this galaxy an outer
bulge seems to be absent or very faint.
The nuclear K band spike is unresolved and is most likely due to emission by
hot dust, as we will discuss in sect.4.1.
The hot dust emission might also affect the H band nuclear peak,
though this is more difficult to test.

Fig.2 shows the H--K color map. To generate such a map we had to center the
K band image relative to the H image. Since
the nucleus might be heavily reddened,
the peak of the H light might be offset with respect to the K peak.
Therefore, we
did {\it not}
 determine the relative position of the two maps by aligning their
peaks. Instead, we used both the cross-correlation of the two maps (excluding
the peaks) and the location of a point source situated north of the nucleus.
The H--K color of an unreddened stellar population does not change
significantly as it evolves; indeed, both spiral and ellipticals show a
fairly narrow distribution of the H--K color, which clusters at 0.22 mags, with
a standard deviation of about 0.1 mag (Frogel
1985, Giovanardi \& Hunt 1996, Hunt et al. 1996).
As a consequence, the H--K excess relative to the standard value
traces regions affected by reddening.
This is not the case for the
Seyfert nucleus, because it has intrinsically redder colors. Moreover,
PSF differences between H and K make color gradients around the nuclear
spike unreliable. However,
that is not a problem for the extended emission, where small
variations of the PSF do not significantly affect the colors.
Fig.2 indicates a color gradient towards the SE.
It is unlikely that this gradient is an instrumental effect due to
imperfect flat fielding, as we found the same result by using different
types of flat fields (night sky, twilight, and even without any flat field).
Such a gradient very likely reflects an increase in the extinction
towards the SE; this variation of the reddening is expected,
as the SE is the near side of the galactic disk and therefore
the dusty screen is expected to be more effective in this region
(Quillen et al. 1995).
However, the most interesting feature is
the dust lane that enters the nuclear region from East toward South and
twists toward the nucleus by 90 degrees at about 80 pc from the center. Such
a dust lane ($A_V\simeq$7.8) can account almost entirely for the interarm
depression near the southern spiral arm observed in the K image (Fig.1a),
though a stellar spiral component is probably also present.
%We will discuss
%further this feature in sect.6.2.

\subsection{Spectroscopic data: emission lines distribution}

In Fig.1a the solid line box indicates the region sampled by 3D 
during the
R=1000 observations. Fig.3a shows the 3D, R=1000, spectrum extracted from an
aperture of $0''.75$ (=15pc) centered on the nucleus.
The spectrum has been rebinned to R$\sim$700 to improve the signal to noise.
Most prominent in this
spectrum are the coronal lines: [SiVI]
(ionization energy 167 eV)\footnote{This is the ionization energy of
the lower ionization stage, i.e. the minimum energy required to obtain the
observed ion.},
[CaVIII] (128 eV)$^4$, and [AlIX] (285 eV)$^4$;
the latter is the first detection in an extragalactic object.
Tab.1 lists the fluxes of the lines detected in the nuclear region.
The continuum rising toward longer wavelengths,
and the stellar features being much shallower than in normal galaxies,
indicates that the non--stellar emission from the active nucleus
contributes
significantly to the continuum, diluting the stellar features.
We extracted line maps by interpolating the underlying 
continuum with a first order fit
in the $\Delta \lambda \sim 0.2 \mu m$
around each line (by excluding regions affected by other
lines and stellar features). An accurate continuum subtraction
is one of the most critical issues, as the continuum slope and shape changes
significantly over the field of view.
Fig.4 shows maps of some of the lines detected in the
K band spectrum. The last contour in each map is at
3$\sigma$ above the noise. The [SiVI] map extends in the direction of the
ionization cone traced by the [OIII] maps, as indicated by means of yellow
dashed
lines (Marconi et al. 1995, Veilleux \& Bland-Hawtorn 1997). In contrast,
the Br$\gamma$ emission is more diffuse and
extends also along the major axis of the
galaxy (P.A.$\sim 25^{\circ}$, see note 5).
The [AlIX] shows the most compact emission.
The nuclear H$_2$ emission is elongated towards the West,
i.e. along the ionization
cone, as [SiVI] does, but at lower surface brightness levels it also
extends towards the NE, along with the Br$\gamma$ emission.
We will discuss these line maps in sections 4 and 5.

\subsection{Spectroscopic data: stellar kinematics}

The dashed rectangle in Fig.1a indicates the region of the nucleus
sampled by the 3D observations at R=2000. Fig.3b shows the
R=2000 nuclear spectrum, extracted from an aperture of $0''.75$.
For comparison, the 3D spectrum of a template star (M3I) is shown
in Fig.3c (note the expanded scale on the {\it y} axis).

We have used such spectra to map the stellar velocity field in the nuclear
region of the Circinus galaxy. Both the galaxy spectra and the stellar
template spectra were continuum subtracted. Then we used two different
methods to determine both the velocity of the stellar system and its velocity
dispersion. We first used a method similar to that described in
Tonry \& Davis (1979):
the cross-correlation of the galaxy spectrum with the template star was
fitted by the autocorrelation function of the star convolved with
a gaussian, whose $\sigma$ and peak velocity were free parameters. The second
method consisted of fitting the galaxy spectrum in real space with a
template stellar spectrum convolved with a gaussian: in this case the
fit parameters were the $\sigma$ of the gaussian profile, the
velocity shift applied
to the stellar spectrum, and a multiplicative factor to fit the depth of the
stellar features. Fitting was performed by using different stellar templates
(K--M giants and supergiants) to monitor possible uncertainties introduced
by a mismatch between the stellar template and the observed galaxy's
 stellar population.
To minimize these mismatch effects, when working in real space, we
separately fitted 
each stellar feature, so that the minimization of the $\chi ^2$
was not limited by differences in the relative depths of the
features between the stellar template and the galaxy spectrum.
Both fitting methods, i.e. fitting of the cross-correlation function and
fitting in real space, gave consistent results.
As the velocity dispersion turns out to be quite low ($\sim$80 km/s), i.e
close to our resolution element, fitting the higher order moments
instead of a simple gaussian (i.e. the kurtosis h$_3$ and h$_4$)
was found not to be provide additional information when
compared to uncertainties.
Errors were estimated by determining the variation of the fit parameters
that decreased the $\chi ^2$ probability by 68\% . Not all the stellar
features in the spectrum could be used: the $^{12}$CO(3--1) band is corrupted
by the [CaVIII] emission line; the $^{13}$CO(2--0) band is affected by a helium
emission line; the $^{12}$CO(4--2)  band is affected by some atmospheric
features or by some unidentified emission line; the NaI doublet is
corrupted by some instrumental effect
 in the nuclear spectrum (see Fig.3b), but could be used
in other regions. Therefore, we used only the $^{12}$CO(2--0) band
and the CaI triplet over the whole field of view, and the NaI doublet
only in the
off nuclear regions (though the latter does not affect the
 results significantly).

Fig.5 shows the map of the stellar peak velocity as derived
by fitting a cubic spline to a grid of 21 points obtained by rebinning the
field of view of 3D, to improve the signal--to--noise of each spectrum.
The errorbars in the outer regions of the map are $\pm 15~km/s$.
Fig.6 shows the stellar peak velocity and velocity dispersion along
the major and minor axis of the galaxy
(P.A. 25$^{\circ}$ and 155$^{\circ}$ respectively)\footnote{
The major axis of the galaxy is subject to some uncertainty. Freeman et al.
(1977) quote P.A. 30$^{\circ}$. The RC3 catalog gives P.A. 40$^{\circ}$.
Elmouttie et al. (1995) give a P.A. $\sim$ 30$^\circ$ based on an AAT
optical image. The near--IR images in Marconi et al. (1995) should be less
affected by extinction and light from
the young stellar population than optical
images are. The major axis of the isophotes of these IR images is at P.A.
$\sim 25 ^{\circ}$. We suspect that the optical images are affected by
patchy extinction in our Galactic plane, especially at the low surface
brightness level. Indeed, the optical images show a weird tilt towards the
East of the northern extremity of the galactic disk, which might be due to
larger Galactic absorption in the northern region. Also, the crowding
of Galactic sources makes it difficult to identify
 of the low surface brightness
isophotes, probably used to measure the inclination of the galaxy.
In this paper we assume the P.A. of the major axis to be that determined
by means of the NIR images of Marconi et al. (1995), i.e. 25$^{\circ}$.
However, results do not change significantly if the actual major axis
is 30$^{\circ}$ as determined by Freeman et al. and Elmouttie et al..}
as measured by means of
the 3D spectra (filled dots).
The stellar velocity dispersion is roughly constant in the
central 100 pc. 
In Fig.6 we also plot, with hollow squares, the peak velocity
determined by using the [NII] emission line data from Oliva et al.
(1994)\footnote{Their long slit was oriented almost in the same direction
as the major axis: P.A. 20$^{\circ}$.}.
The velocity field along the major axis shows a noticeable rotation pattern.

\section{The Seyfert nucleus} \label{seyfert}

\subsection{The nuclear dust emission}

The shallow stellar features in the nuclear K band spectrum, the
slope of the spectrum (Fig.3) and the extremely red colors of the nucleus
indicate that the non--stellar emission from the Seyfert nucleus contributes
significantly to the radiation observed in the nuclear region and that
it dilutes
the stellar features. As discussed by several authors
(Efstathiou \& Rowan-Robinson 1995 EF95, Granato \& Danese 1994 GD94,
Oliva et al. 1995, Thatte
et al. 1997, Pier \& Krolik 1993 PK93, McAlary \& Rieke 1988),
such K band radiation from Seyfert nuclei is very likely emitted by hot
dust, at temperatures close to the sublimation limit ($\sim$ 1500 K), heated
by the intense UV radiation field emitted by the Seyfert source.
As a consequence, the K band images cannot be used to trace
the stellar light in the nuclear region.
The H band nuclear peak is also likely to be
affected by hot dust emission (Thatte et al. 1997, Oliva et al. 1995).

However, the CO stellar bands
provide information about the fraction of the nuclear flux that is stellar.
By assuming that the intrinsic
equivalent width of the CO bands of the stellar
population in the nuclear region
is the same as in the circumnuclear region (R $\sim$ 50--70 pc),
where the contribution of the nuclear
hot dust source is absent, we can then use
the measured EW(CO) on the nucleus to provide
the amount of non-stellar dilution, or the fraction of the observed radiation
that actually comes from stars. As the mass--to--light ratio turns out not
to change significantly within the central $\sim$200 pc (sect.5.2 and Fig.11),
the intrinsic EW of the CO stellar bands is not expected to change much within
the central region. Therefore, the fraction of the stellar light in the nuclear
region can be reliably estimated in the way we just described, with an
accuracy of about 20\% .

In Fig.7 we show the K-band surface brightness
profile around the nuclear peak as derived from the SHARP image (filled dots
and solid line), normalized to its peak intensity, compared
with the fraction of stellar light (hollow circles)
as derived by means of the CO stellar bands. The innermost point
for the stellar light fraction is determined by extracting the 3D spectrum
from an aperture of $0''.3$ (6 pc) from the data of the
night with best seeing ($\sim 0''.3$). The dotted line indicates a fit
to the stellar data using a modified King profile.
It is noticeable that most of the nuclear emission is indeed non--stellar.
If such a nuclear non--stellar source were unresolved, then the observed
K--band profile should be the sum of the shallow stellar profile and
a nuclear PSF, the latter being
normalized to the peak value of the non stellar light (the PSF with this
normalization is plotted with hollow squares connected by a long-dashed line).
The sum of these components is indicated by the short-dashed line
in Fig.7 and, within uncertainties, it
is identical to the observed K profile. Therefore, the non--stellar K--band
source is point like at our resolution, i.e. it has a radius {\it smaller
than 1.5 pc.}

If Circinus were a type 1 Seyfert such a result would not
be surprising, as the size of the region containing dust close to the
sublimation limit is much smaller than 1 pc. Moorwood et al. (1996a) estimate
an intrinsic luminosity of the
Seyfert nucleus of $\sim 10^{10}L_{\odot}$, therefore, by assuming the
grain mixtures given in Rowan-Robinson (1986), we derive a size of the
hot (1500 K) dust region $R_{sub}\simeq 0.03$ pc.
In case of type 2 Seyferts the modelling is more complex. According to the
unified model (Antonucci 1993)
the nuclear hot dust should be partly obscured by the dense circumnuclear
molecular torus. Therefore, the observed 
non--stellar near IR light should be partly associated
with the short--wavelength tail
of the mid-IR emission, and partly to radiation from the nuclear
hot dust transmitted through the torus, the relative contribution of the
two components being dependent on the optical thickness of the torus.
Various authors have modelled the radiative transfer processes through
the putative torus to determine the expected IR emission from type 2 Seyferts
(PK93, GD94, ER95). In the model used by Granato et al. (1997) to fit the
IR spectrum of NGC 1068, the size of the 2.2$\mu$m source is
expected to be $\sim 70 \times R_{sub}$ (at a brightness level of 30\% of the
peak). If we scale the same model to
Circinus, the size of the 2.2$\mu$m source is expected to be $\sim$ 1 pc
in radius, consistent with our upper limit.
Siebenmorgen et al. (1997) have proposed a spherically symmetric
model to specifically
fit the IR emission from Circinus nucleus. However, their model
predicts a size of 4--4.5 pc (in radius) for the emitting region in the
near--IR: such a value is 3 times larger than
our upper limit at 2.2$\mu m$. Therefore, the axisymmetric geometry
expected from the unified model better fits the observational constraints
in terms of size of the emitting region in the near--IR.

In Tab.2
we compare the luminosity
 of the non stellar nuclear K component, as determined from
our data, along with the luminosity in the L$'$ (3.8$\mu m$), M (4.8$\mu m$) and
N(10.3$\mu m$) bands from Siebenmorgen et al. (1997);
the latter data should be little affected by stellar light within the small
aperture listed in Tab.2.
The continuum rises steeply toward longer wavelengths.
When compared to models, either PK93 or GD94 or ER95,
such a steeply rising continuum indicates that the putative torus should
be observed nearly edge--on. More specifically, given that
the light cone in Circinus is quite wide (opening angle $\sim 100^{\circ}$),
ER95's model can reproduce the observed spectral energy distribution
only if the
torus is highly inclined, so that it can significantly self-absorb the near--IR
emitting region.

\subsection{The coronal lines}

As discussed by several authors (Korista \& Ferland 1989, Oliva et al. 1994,
Moorwood et al. 1996a, Oliva 1997, Marconi et al. 1996),
the coronal lines
observed in AGNs are most likely emitted by
gas highly ionized by the hard UV--X ray nuclear radiation.
Our [SiVI] map (Fig.4b), showing the coronal line emission extending
into the narrow line region, supports such a picture. However, the scenario
is more complex.

The [SiVI] (167 eV) coronal line emission map
extends preferentially in the western
direction, up to $\sim$ 50  pc from the nucleus. Instead, the emission region
of the [AlIX] (285 eV) line is much more compact (Fig.4c) and,
if any, it is only
barely resolved. This is better illustrated by Fig.8, where we plot the radial
surface brightness of [SiVI], [AlIX] and of the PSF, within the
sector $-60^{\circ} \le P.A. \le -95^{\circ}$. The [AlIX] radial profile
departs from the PSF only by 1$\sigma$,
while [SiVI] is clearly resolved.
Also, Fig.8 indicates that the
ionization state of the Narrow Line Region changes with radius
in the western direction: at about 20 pc the
[AlIX]/[SiVI] ratio is 2.7 times lower than on the nucleus at a confidence
level of 5$\sigma$. The difference is less pronunced in the northern
part of the ionization cone, where the [SiVI] surface brightness drops
faster and whose radial profile is closer to that of [AlIX].
Possibly, the western part of the radiation cone (P.A. $\sim -90^{\circ}$)
intercepts the dense, circumnuclear molecular clouds in the galactic plane,
thereby lowering
the ionization parameter lower\footnote{Oliva (1997) warns that the ionization
parameter at the inner face of the cloud might be meaningless when dealing
with the coronal line region (CLR),
while a quasi-spherical photoionization
model is more appropriate given the estimated pc-scale size of the CLR.
However, at the distances ($>$20 pc) that we are considering here, the
plane parallel slab model is a good enough approximation for the clouds
in the galactic disk that are illuminated by the AGN continuum.}
and increasing the emissivity of the coronal lines
($\epsilon \propto n^2$).
This would explain the
lower [AlIX]/[SiVI] ratio in this region and the
 enhanced [SiVI] emission with respect to the northern part of the ionization
cone.
Another possibility
is that the ionization mechanism is different in the western extension.
Collisional ionization, due to fast shocks, could contribute in
this region. However, Oliva (1997), Oliva et al. (1994) and Marconi et al.
(1996) show that ionization of the coronal gas
as a consequence of fast shocks is very unlikely in AGNs, and specifically in
Circinus.

Finally, it is interesting to note that the peak of the [SiVI] emission
is slightly shifted, by $0''.15$ (=3pc), with respect to the K band
 nucleus and to the
peak of the Br$\gamma$ and [AlIX] emission. Although
higher resolution data are required to confirm such a
finding, it is worthwhile to compare this result with the ionization structure
expected to occur in the (inner) coronal line region (CLR). As discussed in
Marconi et al. (1994), if the CLR is ``thick'' for Si$^{+5}$, i.e.
the He$^+$ continuum opacity is responsible for determining its ionization
structure, the peak of the [SiVI] emission should occur at the He$^{+2}$
Str\"{o}mgren radius, i.e.
$$
  R_{[SiVI](peak)} = \left[ Q(He^+) \over n_e^2 \alpha_B(He^+)
   \right] ^{1/3}
$$
(assuming a filling factor close to unity). By using
the parameters for the Circinus
CLR given in Moorwood et al. (1996a), i.e. $Q(He^+)\simeq Q(H)/2\simeq
10^{53} phot/sec$ and $n_e\simeq 5000 cm^{-3}$, we estimate
$R_{[SiVI](peak)}\sim $ 5 pc, in agreement with the valuie observed
in the 3D map.
Such gas, responsible for the nuclear coronal emission lines,
must intercept only a fraction of the nuclear solid angle, allowing the
UV--X radiation to ionize gas in the outer light cone.

A more detailed discussion on the coronal emission lines, including the
[CaVIII] line (which requires a careful subtraction of the underlying
stellar features) and their kinematical properties, will be presented in a
forthcoming paper.

\subsection{The molecular gas}

There are three main mechanisms that may be responsible for the emission
of the H$_2$ lines observed in the near IR.
These lines can be excited thermally
in hot gas, either heated by shocks (Hollenbach \& Shull 1977, Draine 1980,
Draine et al. 1983) or by intense X--ray radiation (Lepp \& McCray 1983,
Maloney \& Hollenbach 1996, Tin\'e et al. 1997), or can be emitted under
fluorescent excitation conditions if the UV radiation
field is strong enough (Black \& van Dishoeck 1988, Sternberg 1988).
Near IR spectroscopy of AGN and starburst galaxies has indicated that
UV fluorescence pumping is unlikely to play a major role in most
of these objects
(Moorwood \& Oliva 1990, Sugai et al. 1997), and therefore
thermal excitation must prevail. However, Sternberg \& Dalgarno (1989),
have studied the UV excitation with a more detailed model
and found that
{\it i}) the UV field can contribute to the gas heating and {\it ii})
when the gas densities are higher than 10$^4 cm^{-3}$ collisional
de-excitation of the H$_2$ levels
becomes important, and the emitted spectrum resembles
the thermal case.

The maps of the H$_2$ emission from the Circinus galaxy (Figs.4.d-e)
indicate that the nuclear emission is elongated 10--20 pc
towards the West, i.e.
in the same direction as the X--ray excited [SiVI] line. However, this
is not sufficient evidence to support X--ray heating, as the UV field
in this direction is also
intense. Moreover, at lower surface brightness levels
the H$_2$ emission also extends in the NE direction, i.e. it extends in
the same direction as the Br$\gamma$ emission ($\propto$ UV radiation),
which probably traces star forming activity (sect.5.1).

The H$_2$ line ratios
help to better understand between various excitation mechanisms. The
curves in the diagrams on the left side of Fig.9 show the expected
fluxes of some of the H$_2$ lines, relative to H$_2$(1--0)S(1),
as a function of temperature, for pure thermal excitation.
The dots in the diagrams on the right
side show the same ratios
according to the combined UV pumping+thermal
models of Sternberg \& Dalgarno (1989), as a function of gas density,
for their $\chi$ parameter (proportional to the intensity of the
UV radiation) ranging from 10 to 10$^4$; except for the case of
(1--0)Q(1), for which only the case $\chi=10^2$ is given in their
paper.

The hatched areas indicate the ratios observed on the nucleus,
and their thickness give the $\pm 1\sigma$ range. The lightly--shaded
areas indicate the ratios observed in
an aperture $\sim$ 20 pc NE of the nucleus.
The observed
(2--1)S(1)/(1--0)S(1) ratio clearly rules out pure UV pumping
(the low density regime of Sternberg \& Dalgarno's models) as the excitation
mechanism in both regions. Thermal excitation fits the
data of the H$_2$ nuclear emission, within $\sim 1.2 ~\sigma$,
if the gas has a temperature of about
2500 K. For the UV+thermal models the agreement with the nuclear line ratios
is worse, this is because
in the very high density, quasi-thermal regime the molecular gas reaches
temperatures that are too low ($\sim$1000 K).
However, we cannot exclude that
by varying the parameters in Sternberg \& Dalgarno's model a better
agreement can be obtained.
The similar
western extension of the H$_2$ and
[SiVI] lines, together with these results suggests that the nuclear
H$_2$ lines are mostly emitted thermally in gas heated by the
X--ray, and possibly UV,
 radiation emitted by the AGN. If this were also the case
for other AGNs (e.g. NGC4945, Moorwood et al. 1996b), then the
H$_2$ emission line distribution in active nuclei would not necessarily trace
the distribution of molecular gas, but it 
would also reflect the distribution of the X--ray radiation
field.

In the NE off--nuclear region, where extended Br$\gamma$ emission is seen,
the thermal model reproduces poorly the observed H$_2$ line
intensities,
as different line ratios measure different temperatures. In this region
UV pumping might contribute to the gas excitation. More specifically,
Sternberg \& Dalgarno's model could fit the observed ratios if
the gas density is close to the transition between UV--fluorescence
and thermal regimes, i.e. $\sim 10^{4.5} cm^{-3}$.

\subsection{The nuclear dynamical mass}

As illustrated in Fig.6, within the central 100 pc the bulge is isothermal
within uncertainties.
The smallest aperture over
which we can sample the velocity dispersion is given by
the seeing during the observations, i.e. $\sim 0''.5-0''.6$ (= 10--12 pc).
Though on one night the seeing was as good as $\sim 0''.3$, data from that
night alone do not have a high enough signal--to--noise to sample
the stellar velocity dispersion within the innermost $0''.3$, also due to
higher dilution from the
hot dust component in the central aperture.
Following Kormendy (1988) and Kormendy \& Richstone (1995),
we extracted the spectrum of the nuclear stellar cluster (R$\le$6pc),
cleaned of the bulge contribution, by subtracting
the mean of the spectra at $\pm$17 pc along the minor axis (which have
the same velocity shift as the nucleus).
The resulting stellar spectrum has a velocity dispersion of $55\pm 15~km/s$.
The velocity dispersion is lower in the nuclear region indicating that
we do not detect a nuclear point--like concentration of mass (Binney \&
Tremaine, 1987). By assuming the nuclear stellar cluster to have a
virialized, isotropic distribution and a King core radius
of 10 pc, as derived from Fig.7,
we estimate an upper limit of $4 \times 10^6 M_{\odot}$
 to any compact concentration of mass.
The crucial point, as we will show in sect.5.2,
is that such a nuclear mass is {\it not} associated with a
mass--to--light ratio higher than that of the circumnuclear region,
as would be expected in the case of a
nuclear concentration of dark mass significant with respect to the
stellar mass.

If the AGN is powered by an accreting black hole, then our upper limit
on the black hole mass indicates that the AGN is radiating at
a rate
($10^{10} L_{\odot}$, Moorwood et al. 1996a) that is higher than
0.1 times its
Eddington limit, i.e. L$_{AGN}$/L$_{Edd}>$0.1.
In NGC 1068, another Compton thick Seyfert 2 nucleus,
L$_{AGN}$/L$_{Edd}=$ 0.2 (Greenhill et al. 1996,
Pier et al. 1994). Such high values of the L/L$_{Edd}$ ratio might
be common in these highly obscured AGNs. Indeed, as pointed out by
Granato et al. (1997), the L/L$_{Edd}$ ratio is expected to increase
with the circumnuclear column density (in their model
L/L$_{Edd}\propto N^2_H$).

\section{Star formation and stellar population} \label{starburst}

As discussed in the introduction, the H$\alpha$ and [SII] maps
(Marconi et al. 1995) show evidence for a star forming ring at R = 200
pc, however H$\alpha$ and [SII] emission is also observed
inside the ring.  The 3D Br$\gamma$ map
(Fig.4a) shows evidence for star forming activity in the close
vicinity of the Seyfert nucleus: indeed, the Br$\gamma$ emission extends
outside the ionization cone, along the galaxy major axis, at 20--40 pc
from the Seyfert nucleus.

In this section we discuss the properties of the star formation and
of the stellar population in the nuclear region of Circinus. We
constrain age and luminosity of the stellar population by comparing
the observed Br$\gamma$ equivalent width (EW(Br$\gamma$)) and mass--to--light
ratio with values predicted by the stellar population synthesis code described
in Sternberg \& Kovo (1997)\footnote{This code uses the stellar evolutionary
tracks of Meynet et al. (1994) for solar metallicity, though in Sect.5.3.3
we also consider the case of $Z = 2Z_{\odot}$.}.
We assume
a Salpeter initial mass function between 1 M$_{\odot}$ and 60 M$_{\odot}$,
and a broken power law for lower masses (Miller \& Scalo, 1977).
The star formation history is assumed to be exponentially decaying:
$$SFR \propto exp(-t/t_0).$$

When dealing with the nuclear stellar population (R $<$ 6 pc) we consider
two different models. Model ``A'' assumes the AGN and the starburst to be two
different entities, and that the Br$\gamma$ emission
in the NLR is accounted for
by the AGN. Model ``B'' assumes that the NLR is mostly ionized by a nuclear
starburst (Terlevich's theory).

\subsection{Br$\gamma$ equivalent width}

The ratio between Lyman continuum luminosity ($L^*_{Lyc}$) and
luminosity emitted in the K band ($L^*_K$)\footnote{L$_K$ is defined as the
luminosity emitted in the 1.9-2.5 $\mu m$ spectral range.}
 by the stellar photospheres is a
function of the ratio between the number of massive hot stars and number
of giant-supergiant red stars. Therefore, $L^*_{Lyc}/L^*_K$ provides
information on the star formation history. EW(Br$\gamma$) $\propto
L^*_{Lyc}/L^*_K$ if:
\begin{description}
\item {\it i}) there is no significant contribution from non--stellar sources to
 the ionizing photon flux;
\item {\it ii}) no ionizing photons escape out of the plane of the galaxy;
\item {\it iii}) there is no differential extinction (in the K band) between
 the HII regions and the red giant--supergiant population
 (in this case EW(Br$\gamma$) is independent of foreground extinction);
\item {\it iv}) there is no contribution from non--stellar reprocessed
 light (e.g. hot dust emission) in the K band.
\end{description}

Fig.10a shows the evolution of EW(Br$\gamma$) expected for continuous
star formation ($t_0=10^9 yrs$), an instantaneous
burst ($t_0=10^6 yrs$), and the intermediate case of $t_0= 10^7
yrs$, as derived by means of Sternberg \& Kovo's code, and
assuming that hypotheses {\it i}--{\it iv} are true.

Within the central $\sim$20 pc, the Br$\gamma$ emission
is dominated by the PSF of the innermost part of the
Narrow Line Region (NLR). When considering {\it model A}, regions
contaminated by the NLR must be avoided. The lightly shaded area in Fig.10a
indicates the EW that we observe in the circumnuclear region, at 30--40 pc
radius,
outside the ionization cone (i.e. North and North--East). In this region we
assume negligible deviations from conditions {\it i--iv}.
The age of the burst derived from Fig.10a is naturally quite sensitive to the
assumed length of the burst ($t_0$).

When dealing with {\it model B} (Terlevich's model) the Br$\gamma$ emission
from the NLR must be included in the parameters constraining the stellar
population, i.e we should use the EW(Br$\gamma$) measured on the nucleus
(R $<$ 6 pc). However, as discussed in Sect.4.1, the nuclear stellar light is
significantly diluted by hot dust emission in the K band. Therefore, we
corrected the nuclear EW(Br$\gamma$) by subtracting the continuum contribution
from hot dust. Moreover, the nuclear Lyman continuum flux is known to
escape out of the plane of the galaxy through the ionization cone.
Thus, we corrected for this effect by dividing the EW(Br$\gamma$) by the
covering factor. Moorwood et al. (1996a) estimate the nuclear covering
factor to be 0.05 by comparing the Lyman continuum flux,
as measured by means of the
H$\alpha$ flux from a spatially resolved cloud within the ionization
cone, and the nuclear Br$\alpha$ flux (dereddened according to the nuclear
Balmer decrement).
The hatched area in Fig.10a indicates the nuclear (R $<$ 6 pc) EW(Br$\gamma$)
corrected for hot dust emission and covering factor.
We will discuss the effect of possible differential extinction in Sect.5.3.3.

\subsection{The mass--to--light ratio}

The mass--to--light ratio in the K band, M/L$^*_K$ (see note 9),
of the stellar population
can better constrain both the age of the stellar population and the length
of the burst t$_0$.

The K--band (projected) stellar
luminosity was determined for various apertures by means
of the SHARP K image at radii $1''.5< R <5''$ and by using photometric data
given in Moorwood \& Glass (1984) at larger radii. Similar to what is
required for the EW(Br$\gamma$), we are interested only in the K band light
emitted directly by
stellar photospheres and not in the contribution from reprocessed
light. Therefore, when measuring L$^*_K$
the contribution from the nuclear non stellar spike
was subtracted. At radii $< 1''.5$, we estimated the fraction of
the K--luminosity that is stellar by means of the stellar CO map, as
described in sect.4.1.

The extinction affecting the K--band stellar light
was estimated by means of the H--K color excess with respect to the
unreddened value of 0.22 (sect.3.1).
By assuming the extinction law given in
Koorneef (1983), a foreground screen model gives $A_K=1.4\cdot E_{H-K}$.
The estimated extinction ranges from
an average value of A$_K$=0.3, within 18$''$, to A$_K$=0.63 at $1.''5$.
The latter value is consistent
with that derived by Oliva et al. (1995) using stellar spectra in both
H and K bands, and it is also close to the extinction $A_K = 0.52$ derived
for the innermost ($\sim 1''$) part of the NLR by means of the hydrogen
emission lines (Oliva et al. 1994).
Colors involving shorter wavelength bands
have larger intrinsic scatter amongst field galaxies and are more sensitive to
age and metallicity effects, therefore they are less suitable to derive
the K--band extinction. However, it is possible to perform a consistency
check by comparing the V--K color with that expected for young and
evolved stellar populations. Within the central 18$''$ V--K=4.3
(Moorwood \& Glass 1984), when compared to the average colors of ellipticals
(V--K$\sim$3.3, Frogel et al. 1978) it yields A$_K \sim 0.1$, while if compared
to intrinsic colors expected for young stellar populations
(V--K$\sim 1.7$ for a $10^8~yrs$ old stellar population, Leitherer
\& Heckman 1995) the observed V--K color implies an A$_K$=0.3.
For apertures smaller than $\sim 1''.5$
 the H--K color cannot be used to derive the extinction
$A_K$ because of the contribution from the Seyfert nucleus. Therefore,
we assume the nuclear extinction to be the same as that derived from the
H--K color at $\sim 1''.5$. In sect.5.3.2
 we will discuss the effects of
nuclear extinction (within the nuclear $R<1''.5$) higher than that assumed
here.

The included mass has been calculated by using the [NII] emission line data at
radii $\ge 5''$, and assuming axisymmetric rotation.
At radii $< 2''$ we used the stellar
velocity maps from 3D. We
used the Boltzmann equation for a spherically symmetric space density
distribution whose velocity dispersion is isotropic
(Binney \& Tremaine, 1987).
Initially, we have also assumed the mass tracers to have 
a constant mass--to--light ratio within the central region, in order
to estimate the deprojected volume density of mass tracers from the observed
surface brightness profile (such a volume density $\rho$ enters in the Jeans
equation as a multiplicative factor in the form d$ln(\rho)$/d$ln(r)$).
The latter bit of reasoning might sound circular,
as we will use it to estimate the mass--to--light ratio of the stellar
population; however, it is possible to iterate the procedure and,
within uncertainties, check for self-consistency.
Due to uncertainties both in the measurements and in the model, the errorbars
range from 40\% up to a factor of 2 of the estimated enclosed mass.
The mass derived in such a way is the dynamical mass, which is actually
an upper limit to the stellar mass. Indeed, the gas mass might contribute
significantly to the dynamical mass. There are no CO millimetric
interferometric measurements of the Circinus galaxy, but CO(2--1) data obtained
with a 22$''$ beam (Aalto et al. 1995) indicate that the gas mass enclosed
within the star forming ring is about $6\times 10^8 M_{\odot}$,
if the Galactic conversion factor
is assumed (Kenney \& Young 1989),
i.e. about half of the dynamical mass in the same
region. On the other hand, the galactic conversion factor might not
be appropriate in the central active region of the galaxy (Aalto et al. 1995,
Casoli et al. 1992) and,
therefore, the amount of molecular
gas determined by means of the CO(2--1) line might be
overestimated. Finally, the old stellar bulge population might
contribute significantly to the dynamical mass, thus affecting the
measure of mass--to--light ratio for the starbursting population.

Fig.11 shows the estimated mass--to--light ratio M/L$^*_K$ of the stellar
population within various apertures. Though uncertainties are large,
the M/L$^*_K$ is clearly larger outside the star forming ring than inside it,
indicating a younger stellar population in the inner region.
In the central 30 pc the M/L$^*_K$ ratio increases slightly, probably due to
the mass contribution
from an older stellar population, though it is
still sustantially
 lower than typical M/L$^*_K$ in bulges of normal galaxies and
in ellipticals ($\sim$ 15 and $\sim$ 35 respectively)\footnote{These
values are from Oliva et al. 1995, converted to our units, where we
have deduced L$_K$ from their L$_H$ by assuming H--K=0.22.}
 characterized by a
fairly old stellar population. In particular, we do not find a nuclear peak in
the M/L as would be expected from the presence of a massive dark object.

Fig.10b shows the expected M/L$^*_K$ ratio vs. age of the stellar
population, as determined by using the Sternberg \& Kovo code,
for the same burst
models as in Fig.10a. The dark--shaded area corresponds to
the M/L$^*_K$ determined just inside the star forming ring,
the lightly--shaded area
corresponds to the value determined within 40 pc from the nucleus,
while the hatched area gives the M/L$^*_K$
on the nucleus (R$\le$ 6 pc).
For a given M/L$^*_K$ the age of the stellar population
depends on the burst duration to a lesser extent than the
Br$\gamma$ equivalent width.

\subsection{Starburst models and implications for the
 Starburst--AGN connection}

\subsubsection{Starburst activity within R $<$ 200 pc (model A)}

We initially consider model A, i.e. the AGN and nuclear star formation are two
different entities and the AGN accounts for the Br$\gamma$
emission in the NLR.
By comparing Figs.10a and b, we note that within the central R$\le$40 pc
(lightly shaded areas) the constant star formation
model ($t_0 = 10^9$) does
not fit both observational constraints, i.e. EW(Br$\gamma$) and M/L$^*_K$.
In other words, in case of constant
star formation the burst would
have converted into stars a mass that is
a factor of 100 higher than the observed dynamical
mass. The $\delta$--burst model
($t_0 = 10^6$) also fits poorly both M/L$^*_K$
and EW(Br$\gamma$) constraints,
though modelling of the $\delta$ burst is
more uncertain as it is more sensitive to the discrete sampling of the
stellar evolutionary tracks.
The parameters that best fit the observations within the central 40 pc
are a moderate duration of
the burst, $t_0 \sim 10^7 yrs$, and an age of the stellar
population of $5\times 10^7 yrs$. Similar parameters are also typical
of other starburst nuclei (e.g. Lutz et al. 1996,
Rieke et al. 1993).

If we assume that the same burst model ($t_0=10^7 yrs$) applies to the whole
region within the central 400 pc, then the observed distribution of M/L$^*_K$
(Fig.11) indicates that
the age of the stellar population within this region ranges
from $4\times 10^7 yrs$ to $1.5\times 10^8 yrs$.

It is interesting to note that by including the star forming ring,
i.e. the annulus at 200 pc traced by H$\alpha$
and [SII] (Marconi et al. 1995), the M/L$^*_K$ increases by about a factor
of 2 (Fig.11). This might indicate that such a ring has undergone
a more recent burst and that it has not yet entered in the red-supergiants
phase. Circinus
could be experiencing an outward propagating starburst: the nuclear
region (R $\le$ 100--150 pc)
experienced a burst about $0.4-1.5\times 10^8 yrs$ ago,
and its near--IR continuum is now dominated by red supergiants,
while the starburst
activity has propagated to R$\sim$200 pc where it is currently forming
young hot stars.
Such a phenomenology is common to other starburst galaxies, such as
M82 and IC342, where evidence has been found for a nuclear stellar
population dominated by red supergiants few times $10^7$ yrs old, while a
circumnuclear ring is undergoing a more recent burst that is producing
young hot stars (Rieke et al. 1980, Satyapal et al. 1997,
F\"{o}rster et al. 1996, B\"{o}ker et al. 1996, 1997).

Fig.10c shows the evolution of the bolometric--to--K band luminosity
ratio, $L^*_{bol}/L^*_K$, as a function of age, for
the same models as in Figs.10a and b. Given an average age of $\sim
10^8 yrs$ for 
the stellar population inside R=200 pc, we get
$L^*_{bol}/L^*_K \simeq 50$ for the $t_0 = 10^7 yrs$ model. Note that the
derived $L^*_{bol}/L^*_K$ ratio is not very sensitive to the adopted burst model
($t_0$), if the M/L$^*_K$ is used as the age indicator, as the various models
in Fig.10b and Fig.10c have opposite trends versus age.
In the extreme case of constant star formation (which
we ruled out) the $L^*_{bol}/L^*_K$ ratio increases only by a factor of 1.5,
while a $\delta$--burst would give the same $L^*_{bol}/L^*_K$ ratio as the
$t_0 = 10^7 yrs$ model.

The stellar bolometric luminosity can be derived from the observed
K--band stellar luminosity by using the following relation:
\begin{equation}
L^*_{bol} = \left( \frac {L^*_{bol}}{L^*_K} \right) _{model} \times
L^*_K(obs)
\end{equation}

where $L^*_K(obs)$ is the observed K--band stellar luminosity corrected
for extinction.

Given the K band luminosity within R $\le$ 200 pc, the total bolometric
luminosity emitted by the stellar population within the same aperture is
$L^*_{bol}\sim 1.1\times 10^{10} L_{\odot}$. This luminosity has to be compared
with the intrinsic AGN luminosity, $10^{10} L_{\odot}$ (Moorwood et al. 1996a),
and with the bolometric luminosity of the galaxy,
$\sim 1.7\times 10^{10} L_{\odot}$
(corrected for Galactic absorption,
but not for internal extinction)\footnote{The IR luminosity from IRAS data
is $1.2\times 10^{10} L_{\odot}$ (Siebenmorgen et al. 1997). By using the
photometric data in Moorwood \& Glass (1983) along with the B$_T$
in the RC3 catalog, corrected for a Galactic extinction $A_V=1.5$,
we estimate a luminosity of $\sim 5 \times 10^{9} L_{\odot}$ in the K to U
wavelength range.}, i.e. Seyfert nucleus and starburst contribute to a similar
extent to the total bolometric luminosity.
Tab.3 summarizes the luminosity budget of the various sources in Circinus.
It is not surprising that the sum of the
starburst and AGN luminosities is slightly larger than the observed
bolometric luminosity: in
this edge--on system a fraction of the light escapes perpendicular to the
galactic plane and, therefore, is not reprocessed into IR radiation
nor is directly observed; in particular, a significant fraction of the
radiation from the AGN is probably lost through its light cones.
From the energetics point of view Circinus is similar to
Mrk 477, the most powerful Seyfert 2 in the local Universe,
for which Heckman et al. (1997) estimate that the nuclear (few 100 pc)
starburst and the Seyfert nucleus provide most of the
total luminosity of the galaxy, in about equal proportions.
However, the starburst in Mrk 477 is at a much
earlier stage (6 Myr) than the starburst in Circinus.

\subsubsection{Starburst activity within R $<$ 6 pc (model A)}

In the innermost 12 pc the Br$\gamma$ emission is affected by the
emission from the NLR. If we assume that the intrinsic EW(Br$\gamma$)
of the starburst within the central 12 pc is the same
as that measured at $\sim$ 30--40 pc, then the nuclear M/L$^*_K$ ratio
implies an age of the nuclear star cluster of about $7(\pm 3) \times 10^7$ yrs,
and a burst length $t_0$ = $10^7 yrs$.

The $L^*_{bol}/L^*_K$ derived for this starburst model (Fig.10c) implies
a bolometric luminosity (eq.1) of the
nuclear stellar population of L$^*_{bol}$(R$\le$6pc) = $2(\pm 0.6)\times 10^8
L_{\odot}$. Therefore, the nuclear stellar luminosity accounts for only
$\sim 2$\% of the Seyfert luminosity estimated by Moorwood et al. (1996a).

As discussed in the introduction, David (1987b),
Norman \& Scoville (1988) and Murphy et al. (1991) proposed models
that ascribe the AGN fuelling to mass 
loss from post--main sequence stars. The gas ejected from post--main
sequence red giants and supergiants is kinematically hot, as its velocity
dispersion reflects the velocity dispersion of bulge stars ($\sim 80~
km/s$ for Circinus, Fig.6), and therefore it is characterized by
the large turbulent viscosity that is required for the gas to inflow.
More specifically,
Norman \& Scoville's (1988) model
 predicts the ratio between the luminosity of the active nucleus
and the luminosity 
of the evolving stellar cluster, within the central 10 pc,
to be $\sim$50 when the
nuclear cluster is 70 Myrs old.
This number is in fair agreement with the ratio between the intrinsic
Seyfert luminosity and the luminosity
derived for the star cluster in the nuclear 12 pc of Circinus:
$L_{AGN}/L_*(R\le 6pc)\approx 46$.
This result does not necessarily prove these models to be correct,
but it is the first time that observational
data provide a consistency check for the
theory of AGN fuelling through stellar mass loss.
Finally, we note that Thatte et al. (1997) and Oliva et al. (1995) have found,
though on larger scales, that other Seyfert nuclei are characterized by
similar moderately young stellar populations (few $10^8$ yrs). It is
tempting to speculate that such intermediate age stellar populations are
closely linked to Seyfert activity in the way we just depicted.

The working hypothesis of this model is that the extinction in the K band
toward the nucleus is the same as that estimated by means of
the H-K color at 30 pc
from the nucleus (i.e. $A_K(nuc)=0.63~mag$, sect.5.2). However, the
nucleus might be affected by an extinction higher than in the circumnuclear
30 pc. Correcting the nuclear K band luminosity for higher extinction
 results in a
lower M/L$^*_K$ ratio, i.e. a younger stellar population.
The M/L$^*_K$ ratio cannot be lower than the minimum value predicted by the
model (Fig.10b), this constrains the maximum possible excess
nuclear extinction to be
$\Delta A = 21.5 ~mag$.
If L$^*_K$(nuc) is corrected for excess nuclear extinction, then
the estimated bolometric luminosity of the stellar population, L$^*_{bol}$,
increases both because of the increased L$^*_{bol}$/L$^*_K$ derived
from the model (as the
lower M/L$^*_K$ indicates a younger population) and because of the
increased L$^*_K$ after correcting for extinction (eq.1).
Fig.12 indicates that if the extinction toward the nucleus is higher than
what we derived from the circumnuclear region, the nuclear stellar cluster
could be younger (as young as $10^7 yrs$) and more luminous (as luminous
as L$^*_{bol} = 2\times 10^9 L_{\odot}$). With regard to the implications
for Norman \& Scoville's model, a younger star cluster would provide an
even larger mass loss available to fuel the accreting black hole.

\subsubsection{Starburst activity within R $<$ 6 pc (model B)}

In model B (Terlevich's model)
most of the luminosity of the active nucleus is due
to starburst activity. In this case, as discussed in sect.5.1,
the nuclear starburst
also contributes to the ionization of the Narrow Line Region. Therefore,
we should use the nuclear EW(Br$\gamma$), corrected as
described in sect.5.1 (hatched area in Fig.10a),
to model the nuclear stellar population.
A $5\times 10^6 yrs$ old $\delta$--burst fits both
the EW(Br$\gamma$) and M/L$^*_K$ ratio.
The age derived for the nuclear star cluster is close to that expected in
Terlevich's model during the Sy2 phase. However, the derived bolometric
luminosity
is $\sim 20\times 10^8 L_{\odot}$, which is only 20\% of the luminosity of the
Seyfert nucleus\footnote{We should
mention that the starburst models were also run by
assuming the metallicity to be twice the solar value, to account
for Terlevich's claim that AGN--like starbursts should be evolving
in a metal rich environment, and the results were found not to change
significantly.}.

However, the main problem is not the luminosity deficit, but the Spectral
Energy Distribution (SED). Our starburst model predicts a SED peaking
close to the Lyman edge, while the AGN radiates most of its energy
around 100 eV and in the hard X--rays (Moorwood et al. 1996a).
On the other hand, our model does not take into account the emission from
SNR radiative shocks, expected to play a role at high energies
in Terlevich's model.

There are other factors that might affect our estimate of the bolometric
luminosity of the nuclear star cluster.
If the foreground extinction is not uniform, but the nuclear star cluster
is more obscured than the Narrow Line Region (differential extinction),
then the observed EW(Br$\gamma$) is higher than that expected from
L$^*_{Lyc}$/L$^*_K$. In this case, the extinction-corrected
EW(Br$\gamma$) is lower than the observed value and the derived age of
the burst is older, as shown in Fig.13.
The corrected M/L$^*_K$ ratio cannot be lower
than that expected by the model, and this constrains the maximum differential
extinction to be A$_V$(stars)--A$_V$(NLR) $<$ 30 mag. The derived bolometric
luminosity changes as a consequence of two competitive factors: {\it i)}
L$^*_K$(corrected) increases, but {\it ii)} $(L^*_{bol}/L^*_K)_{model}$
decreases as a consequence of the older age. The net behaviour is an
increase of L$^*_{bol}$, as shown in Fig.13. The corrected L$^*_{bol}$
could be as high as $7\times 10^9 L_{\odot}$, for a differential
extinction of 30 mags.

Moorwood et al. (1996a) estimate the Lyman continuum luminosity by using
the H$\alpha$ flux from a ``resolved'' cloud in the NLR, whose {\it
projected} distance from the nucleus is known. However, there are a
few factors which lead to an underestimate of the real L$_{Lyc}$:
{\it i)} a fraction of the H$\alpha$ flux from the cloud might be
lost as a consequence of dust extinction within the cloud, {\it ii)}
a fraction of the ionizing photon flux might be absorbed by dust
in front of or within the cloud, {\it iii)}
the distance of the cloud from the nucleus might be larger than the
observed projection, {\it iv)} the cloud might be made up of subclumps or
filaments, i.e. the effective area of the cloud might be lower than observed.
If the real L$_{Lyc}$ is higher
than that estimated by Moorwood et al. (1996a),
then in sect.5.1 we have undercorrected the nuclear
EW(Br$\gamma$) when accounting for the covering factor. Therefore,
correcting the L$_{Lyc}$ for effects {\it i--iv}
makes the nuclear EW(Br$\gamma$) larger and makes
the derived age for the cluster younger, as shown in Fig.14. The
maximum value of the EW(Br$\gamma$) predicted by the model imposes a maximum
correction to the Lyman continuum luminosity:
L$_{Lyc}$(corrected)/L$_{Lyc}$(observed) $<$ 13.
Correcting L$_{Lyc}$ affects both the derived L$^*_{bol}$ (because of the
younger age of the cluster) and the bolometric luminosity of the AGN
(because the total L$_{Lyc}$ is the normalizing
factor of the SED derived by Moorwood et al. 1996a). The net effect
is a decrease of the ratio L$^*_{bol}$/L$_{bol}$(AGN). The most
extreme correction
would imply a very young nuclear cluster ($\sim 10^6 yrs$) which radiates about
8\% of the luminosity of the Seyfert nucleus.

Summarizing, if we assume that a compact nuclear starburst is responsible
for most of the Br$\gamma$ emission of the Seyfert nucleus, the starburst must
be as young as $5\times 10^6 yrs$, in fair agreement with the expectation of
Terlevich's model, but the bolometric luminosity of the starburst is
only 20\% of the observed luminosity of the active nucleus. However,
if the nucleus is affected by differential extinction the luminosity derived
for the nuclear starburst could be higher by up to a factoor of 4.
Yet, the main
problem of fitting the observed SED in Circinus at high energies with
the softer emission expected from the nuclear starburst remains unsolved.

Finally, it is interesting to note that Veilleux \& Bland-Hawthorn (1997)
estimated that the filamentary ejecta, extending radially from the
nucleus of Circinus, were probably expelled by an explosive
nuclear event that occurred a few Myr ago. Such a time scale fits with
the age of the nuclear burst estimated in this section.

\section{Summary} \label{sum}

We have studied the nuclear region of the Circinus galaxy by means
of near IR images which have an angular resolution
$\simeq 0''.15$ (= 3 pc as projected on the source), and K--band
integral field spectroscopy with an angular resolution $\sim 0''.5$
and spectral resolution of 1000 and 2000.

The K--band image reveals a nuclear non--stellar source
that is unresolved, i.e. whose size is smaller than 1.5 pc in radius.
Such nuclear non--stellar light is most likely emitted by dust heated by
the Seyfert nucleus. When compared to models of the
IR emission from obscuring torii, our upper limit fits the expected size
of the K--band emission in Sy2s.

The Coronal Line Region traced by the [SiVI] line extends for
$\approx$ 50 pc into the Narrow Line Region, but the higher excitation
[AlIX] emission line is much more compact. The variation of the
[AlIX]/[SiVI] ratio in the light cone might indicate that the
ionization parameter of the NLR clouds decreases radially
due to photoionization of high density clouds in the galactic plane.

The nuclear emission from the H$_2$ lines observed in the K-band extends
in the same direction as the [SiVI] emission. The H$_2$ line ratios
indicate that such lines are emitted by thermally excited gas, probably
heated by the same X--ray radiation that excites [SiVI]. At lower
surface brightness levels the H$_2$ emission extends along the plane of the
galaxy: in these regions UV--fluorescence might contribute to the excitation
mechanism.

The stellar velocity field traced by the stellar features in the K--band
does not show evidence for a point--like mass concentration in the
nuclear region. Also, there are no hints of an increase in the
mass--to--light ratio toward the nucleus. We set an upper limit of $4\times
10^6 M_{\odot}$ for the mass of a putative black hole.
As a consequence, the L$_{AGN}$/L$_{Edd}$ ratio must be larger than 0.1.

The Br$\gamma$ map indicates evidence for ongoing star forming activity
within a few tens of pc from the active nucleus. The mass--to--light ratio,
from the 100 pc scale to the innermost 10 pc, is quite low and indicates
that the stellar population is relatively young, with an age ranging
between $4\times 10^7$ and $1.5\times 10^8 yrs$. There are indications
of an outward propagation of the starburst.  The bolometric luminosity
derived for the young stellar population inside the central 400 pc is
comparable to the (intrinsic) luminosity of the active nucleus, and both
of them together contribute most of the bolometric luminosity of the galaxy.
The stellar population within R $< 6$ pc  has an age of $7\times 10^7$ yrs
and contributes only 2\% of the intrinsic luminosity of the active nucleus.

We have discussed models that ascribe AGN fuelling to mass loss from
a young nuclear star cluster in the light of our observational constraints.
Our results fit predictions of these models in terms of expected
 nuclear stellar luminosity relative to the AGN luminosity.

We have also considered the case
of a nuclear starburst
contributing significantly to the Br$\gamma$ emission
in the Narrow Line Region (Terlevich's hypothesis). In this case
the nuclear ($\sim$ 10 pc) stellar population
could be as young as $5\times 10^6 yrs$,
but the implied bolometric luminosity
of the nuclear star cluster would be only 20\% of the intrinsic
luminosity of the active nucleus. However, differential
extinction effects could lead to an underestimate of the
luminosity of the nuclear star cluster by up to a factor of 4.

\acknowledgments

We appreciate comments and helpful discussions with A. Moorwood,
T. Storchi-Bergmann and G.L.Granato. We are grateful to S. Anders,
A. Eckart, J. Gallimore and H. Kroker for help during the observations.
We thank A. Sternberg for making available his code.
R.M. acknowledge the support of the ASI grant ASI-95-RS-120.

\clearpage

\figcaption{
$a)$ K--band SHARP image, smoothed with a 3 pixel (=$0.15''$)
 FWHM gaussian.
Contours give Log(F$_{\lambda}$), in units of
$10^{-24}~W~cm^{-2}\mu m^{-1} arcsec^{-1}$, at the following levels:
0.6, 0.75, 0.8, 0.9, 1.0, 1.1, 1.2, 1.3, 1.4, 1.5, 1.6, 1.7, 1.8.
The green solid--line and dashed--line boxes show the regions
observed with 3D at a spectral resolution of R=1000 and R=2000 respectively.
$b)$ Enlargement of the central $3''$ of the unsmoothed K--band SHARP image.
Contours are in the same units as in Fig.1$a$ at the levels:
1.2, 1.3, 1.4, 1.5, 1.6, 1.8, 2.0, 2.2, 2.4.}

\figcaption{
H--K color map, smoothed as Fig.1$a$. Contours are at 0.25, 0.35,
0.45, 0.55, 0.65, 0.75 mags. Regions affected by low
signal--to--noise have been blanked.}

\figcaption{
$a)$ Nuclear spectrum of Circinus, extracted from an aperture of 0.75$''$,
 rebinned to a spectral resolution of 700. Labels marked with a ``*'' show the
position of absorption stellar features, while other labels identify emission
lines. $b)$ The long wavelength part of
the nuclear spectrum (same aperture as $a$) at a spectral resolution of 2000.
$c)$ Spectrum of one of the template stars (HD94613, shifted to the
redshift of Circinus) used to fit the
profile of the stellar features in Circinus' spectra.}

\figcaption{
Maps of emission lines in the K band spectrum. The cross marks the position
of the continuum peak.
The last contour in each map is at 3$\sigma$ above the noise. $a)$ Br$\gamma$;
contours are at 80\% ,60\% , 40\% , 30\% , 20\%, 15\% and 10\% of the peak.
$b)$ [SiVI] (the yellow dashed lines indicate the position of the ionization
cone traced by the [OIII] line maps from Marconi et al. 1995 and
Veilleux \& Bland-Hawthorn 1997);
contours are at 90\% ,60\% , 40\% , 30\% , 20\%, 15\% , 7.5\%  and 5\%
of the peak.
$c)$ [AlIX];
contours are at 80\% , 60\% , 40\% , 30\% , 20\%, 17\% and  12\% 
of the peak.
$d)$ H$_2$(1--0)S(1);
contours are at 80\% , 60\% , 40\% , 30\% , 20\%, 15\%
of the peak.
$e)$ H$_2$(1--0)S(2);
contours are at 80\% , 60\% , 50\% , 40\% , 28\% 
of the peak.
In $d$ and $e$ two areas affected by bad pixels have been blanked.}

\figcaption{
Map of the stellar velocity
field as measured by means of the IR stellar
features in the 3D spectra.
Contour levels are spaced by 10 km/s, from 224 to 294 km/s.
Errorbars in the outer parts of the map are $\pm 15$ km/s.}

\figcaption{
Stellar
peak velocity and velocity dispersion as measured by means of the stellar
features in the 3D spectra (filled dots).
Hollow squares indicate the velocity
derived by means of the [NII] emission line data in Oliva et al. (1994).}

\figcaption{
Radial profile of the nuclear K--band surface brightness
 as derived from the SHARP
image (filled circles and solid line). Hollow circles indicate the fraction of
the K--band light emitted by stars as determined by means of the 3D maps
of the CO stellar bands. The dotted line indicates a fit of
the stellar fluxes by using a modified King profile. The hollow squares
connected by a long dashed line indicated the PSF profile normalized
to the peak of the non stellar light. The short-dashed line is the sum
of the stellar component (dotted line) and of the PSF (long-dashed line).}

\figcaption{
Radial profile of the surface brightness of the [SiVI] and [AlIX] emission
lines, compared to the PSF, in the sector $-60^{\circ}\le P.A. \le
-95^{\circ}$.}

\figcaption{
Flux of some of the H$_2$ lines detected in the K band relative
to the H$_2$(1--0)S(1) line. The hatched
areas indicate the values observed on the nucleus ($0.6''$ aperture),
while the lightly--shaded areas indicate the ratios observed in an aperture
at about 20 pc NE of the nucleus. The thickness of the shaded areas
gives the $\pm \sigma$ range. The curves in the diagrams on the left side
indicate the ratios expected in case of emission from thermally excited
gas, as a function of temperature. The dots in the diagrams on the right
side give the ratios predicted by Sternberg \& Dalgarno's models (see text),
as a function of gas density, for their parameter $\chi$ (that is
proportional to the UV radiation field) ranging from 10 to 10$^4$;
except for the (1--0)Q(1) line where only the $\chi = 10^2$ case is given.}

\figcaption{
$a)$ Expected quivalent width of the Br$\gamma$ versus age of a starburst
in case of a $\delta$ burst (dotted line), a burst duration of
$10^7 yrs$ (solid line) and continuous star formation (dashed line).
The lightly--shaded area indicates the observed range of EW(Br$\gamma$)
at 30--40 pc from the nucleus, outside the ionization cone.
The hatched area indicates the nuclear
EW(Br$\gamma$) after correcting the Br$\gamma$ flux for the covering factor
of the NLR and the K band light for contamination by hot dust.
$b)$ Expected mass--to--light ratio for the same models as in $a$.
The darkly--shaded area gives the M/L$^*_K$ determined inside the star forming
ring, the lightly--shaded area gives the same ratio inside R = 40 pc,
while the hatched area gives the M/L$_K$ on the nucleus (R$\le$ 6pc).
The solid--line and dashed--line boxes indicate the range of values observed
in bulges of spirals and in ellipticals (from data in Oliva et al. 1995).
$c)$ Expected bolometric--to--K band luminosity for the same models
as in $a$.}

\figcaption{
Mass--to--light ratio within various regions centered on the nucleus.}

\figcaption{
Effects of correcting the nuclear K--band light for excess nuclear extinction
with respect to the circumnuclear region, in model A.
{\it Top}, effect on the age derived for the nuclear star cluster.
{\it Bottom}, effect on the bolometric luminosity derived for the nuclear
star cluster. See text about estimates of the maximum excess extinction.}

\figcaption{
Effects of correcting the nuclear K--band light and EW(Br$\gamma$) for
differential extinction between nuclear stellar population and
Narrow Line Region, in model B.
{\it Top}, effect on the age derived for the nuclear star cluster.
{\it Bottom}, effect on the bolometric luminosity derived for the nuclear
star cluster. See text about estimates of the maximum differential
extinction.}

\figcaption{
Effects of correcting the nuclear EW(Br$\gamma$) for asborption of
UV and H$\alpha$ photons in the NLR and other factors possibly responsible
for reducing the $L_{Lyc}$ derived from the observations (see text),
in model B.
{\it Top}, effect on the age derived for the nuclear star cluster.
{\it Bottom}, effect on the ratio between the bolometric luminosity derived for
the nuclear star cluster and the bolometric luminosity of the AGN.
See text about estimates of the maximum L$_{Lyc}$ correction.}

\clearpage

\begin{deluxetable}{lcc}
\small
\tablecaption{Nuclear emission lines (aperture = $0.75''$)}
\tablewidth{0pt}
\tablehead{
\colhead{Line} & \colhead{$\lambda ^a$} & \colhead{Flux$^b$} \\
\colhead{}   & \colhead{($\mu$m)}  &
 \colhead{($10^{-15}erg~cm^{-2}s^{-1}$)} \\ }
\startdata
H$_2$(1--0)S(3) & 1.9576 & 14.4$\pm$5 \nl
$[$SiVI$]$          & 1.9635 & 70.2$\pm$5 \nl
H$_2$(1--0)S(2) & 2.0332 & 6.09$\pm$1.2 \nl
$[$AlIX$]$          & 2.043  & 15.6$\pm$1.3\nl
HeI             & 2.0581 & 5.5$\pm$0.7\nl
H$_2$(1--0)S(1) & 2.1218 & 11.5$\pm$0.6\nl
Br$\gamma$      & 2.1655 & 15.5$\pm$0.6\nl
HeII            & 2.1885 & 2.4$\pm$0.6\nl
H$_2$(1--0)S(0) & 2.2235 & 2.07$\pm$0.46\nl
H$_2$(2--1)S(1) & 2.2471 & 1.13$\pm$0.37\nl
$[$CaVIII$]$        & 2.3213 & 47.5$\pm$5 \nl
H$_2$(1--0)Q(1) & 2.4066 &  9.5$\pm$1.9\nl
\enddata
\tablenotetext{a}{Wavelengths are in the rest frame}
\tablenotetext{b}{Fluxes are corrected for a Galactic
extinction of $A_V=1.5~mag$, but uncorrected for internal reddenning.}
\end{deluxetable}

\begin{deluxetable}{lcc}
\small
\tablecaption{Luminosities and sizes of the nuclear non stellar source
at various wavelengths in the near--to--mid IR spectral region.}
\tablewidth{0pt}
\tablehead{
\colhead{$\lambda $ ($\mu$m)} &
\colhead{$\nu$L$_{\nu}$ ($\times 10^7 L_{\odot}$)} &
\colhead{Radius (pc)}
\\ }
\startdata
2.2 (K)$^a$ & 1.5 & $<$1.5 \nl
3.8 (L$'$)$^b$ & 30 & 3 \nl
4.8 (M)$^b$ & 56 & 3 \nl
10.3 (N)$^b$ & 90 & 13 \nl
\enddata
\tablenotetext{a}{From the SHARP image,
corrected for the contribution of the stellar
light as derived from the stellar CO map (Fig.6)}
\tablenotetext{b}{From
Siebenmorgen et al. (1997).}
\end{deluxetable}

\begin{deluxetable}{lccc}
\small
\tablewidth{0pt}
\tablecaption{
Starburst models and
luminosity budget of the various sources in Circinus.}
\tablehead{
\colhead{Source} &
\colhead{Luminosity} &
\colhead{Age} &
\colhead{Burst duration}
\\ }
\startdata
Observed$^a$  & $1.7\times 10^{10} L_{\odot}$ & & \nl
AGN intrinsic$^b$  & $10^{10} L_{\odot}$  & & \nl
Starburst (R$<$200pc)  &
   $1.1\times 10^{10} L_{\odot}$ & $10^8yrs$ & $10^7 yrs$ \nl
Starburst (R$<$6pc), mod. A  &
   $2 \times 10^8 L_{\odot}$ & $7\times 10^7 yrs$ & $10^7 yrs$ \nl
Starburst (R$<$6pc), mod. B  &
   $2 \times 10^9 L_{\odot}$ & $5\times 10^6 yrs$ & $10^6 yrs$ \nl
\enddata
\tablenotetext{a}{Bolometric luminosity corrected
for Galactic extinction, but not for internal reddening
(see note 10).}
\tablenotetext{b}{From Moorwood et al. (1996a).}
\end{deluxetable}

\begin{figure}
\figurenum{3}
\epsscale{1}
\plotone{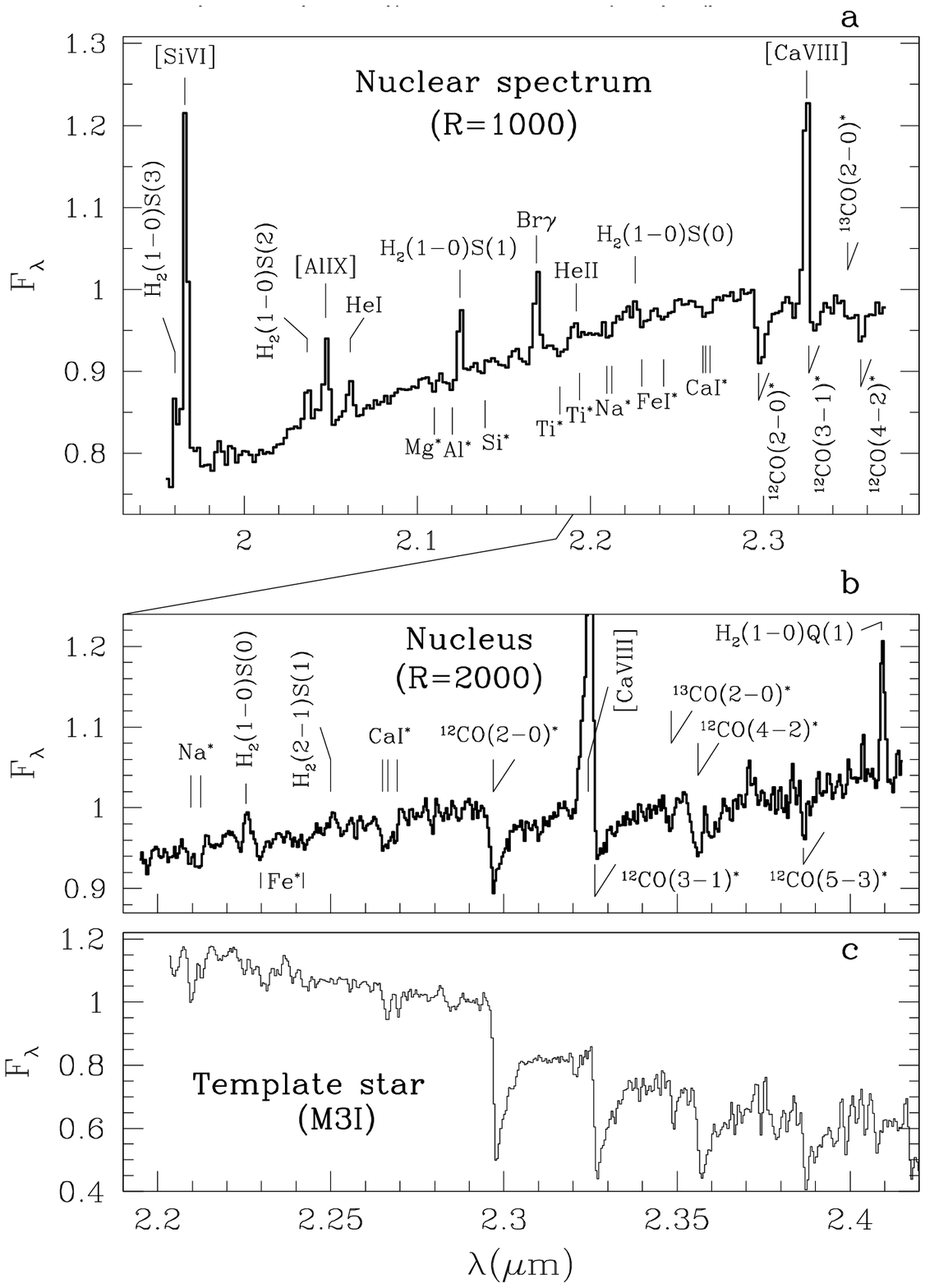}
\caption{}
\end{figure}

\begin{figure}
\figurenum{5}
\epsscale{1}
\plotone{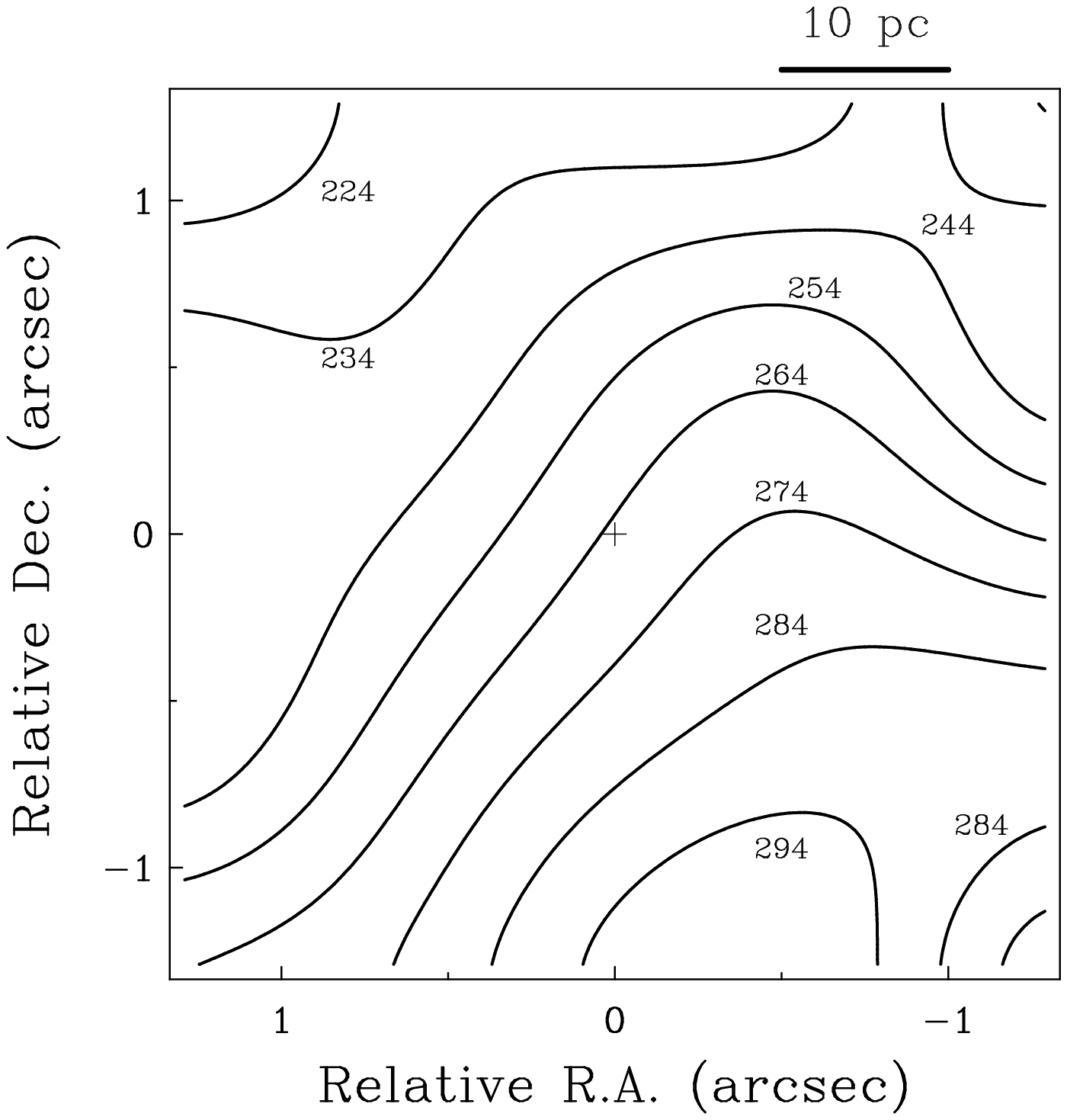}
\caption{}
\end{figure}

\begin{figure}
\figurenum{6}
\epsscale{1}
\plotone{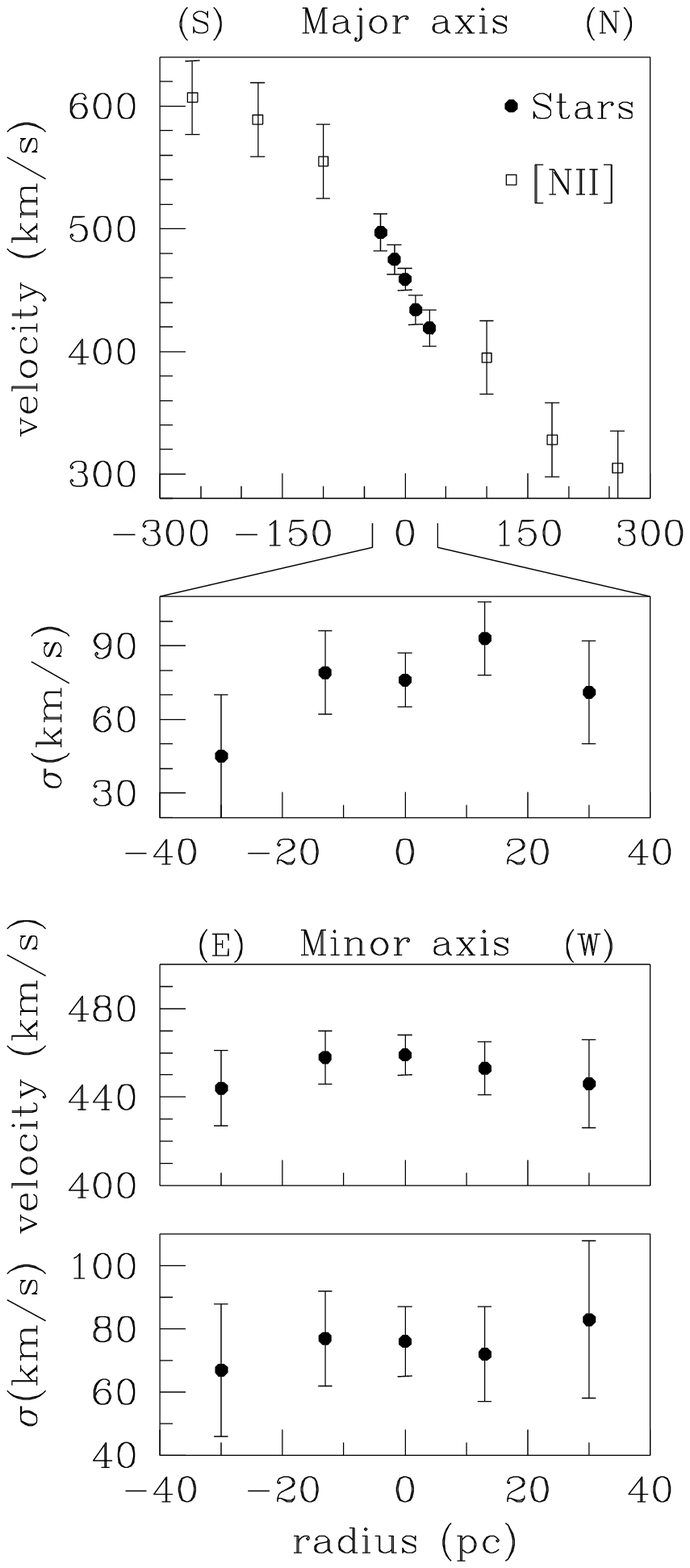}
\caption{}
\end{figure}

\begin{figure}
\figurenum{7}
\epsscale{1}
\plotone{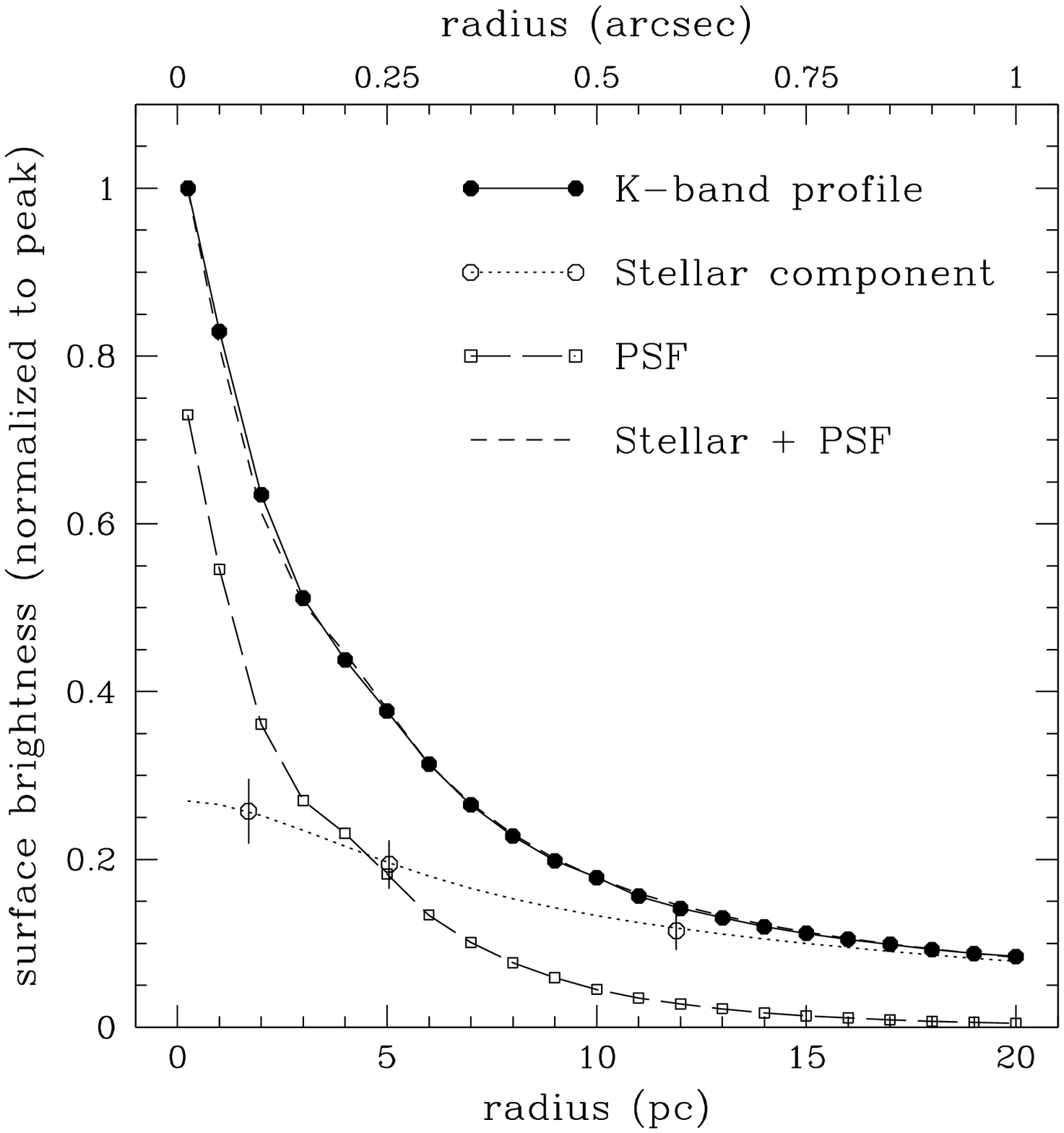}
\caption{}
\end{figure}

\begin{figure}
\figurenum{8}
\epsscale{1}
\plotone{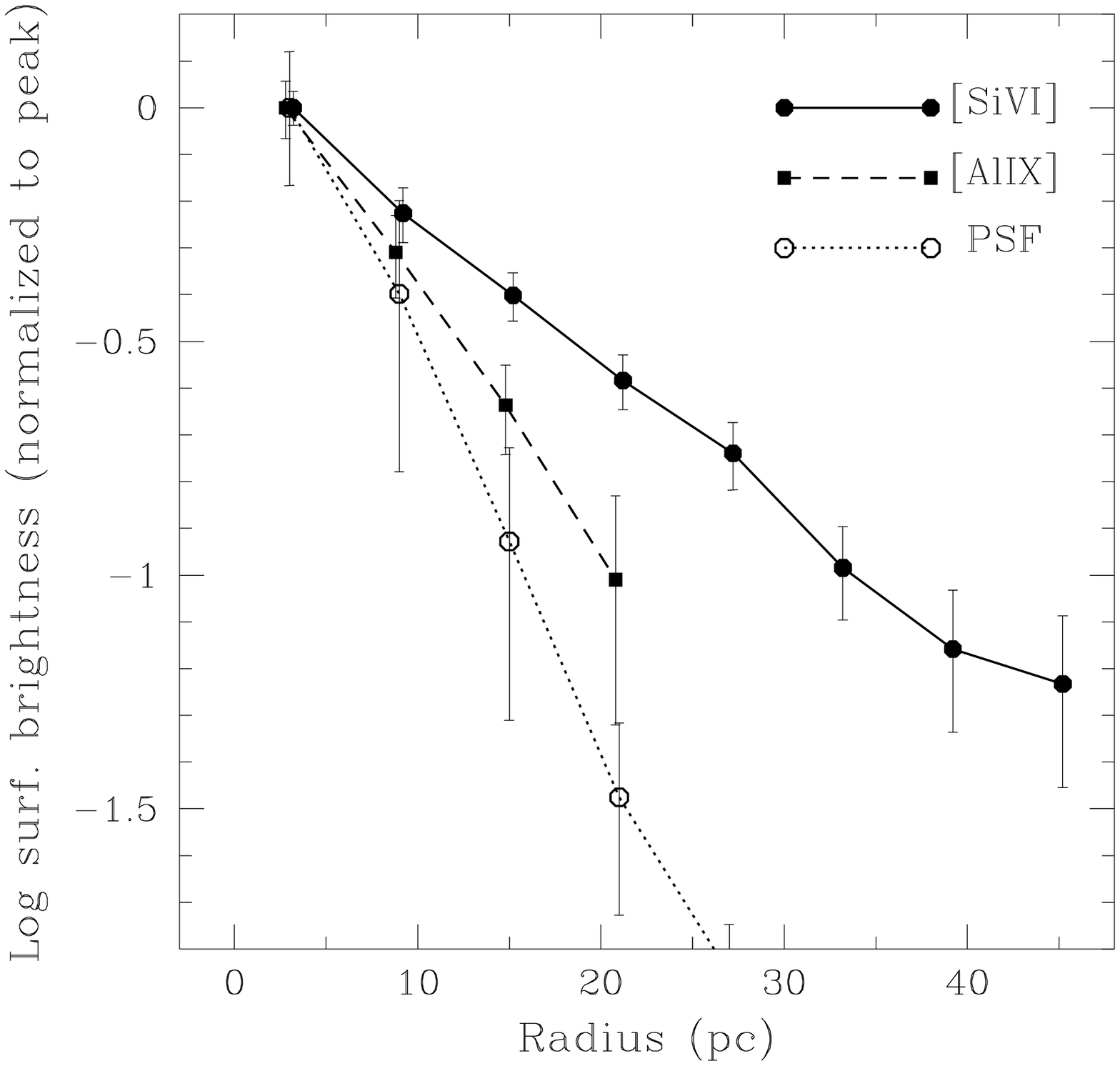}
\caption{}
\end{figure}

\begin{figure}
\figurenum{9}
\epsscale{1}
\plotone{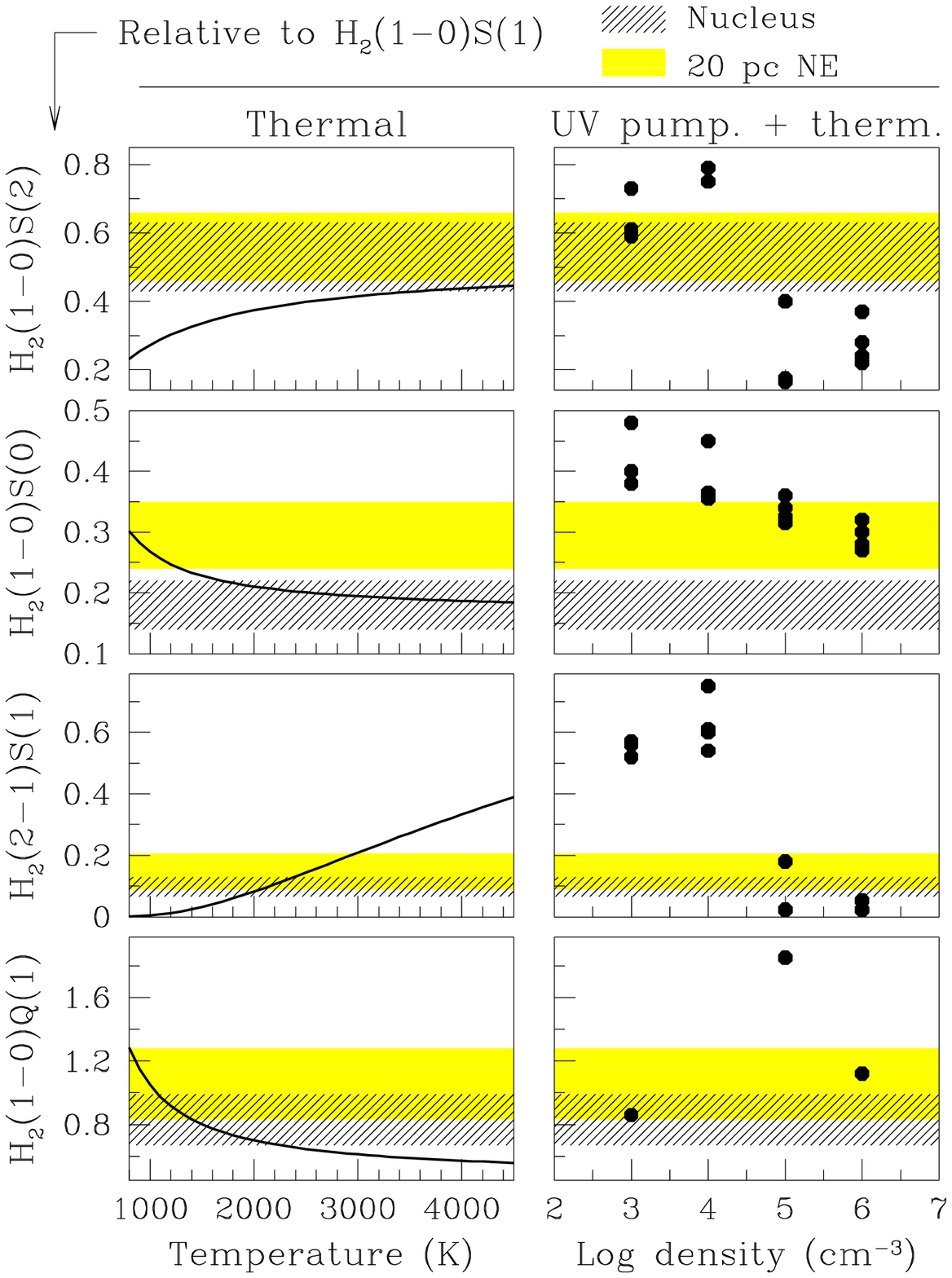}
\caption{}
\end{figure}

\begin{figure}
\figurenum{10}
\epsscale{1}
\plotone{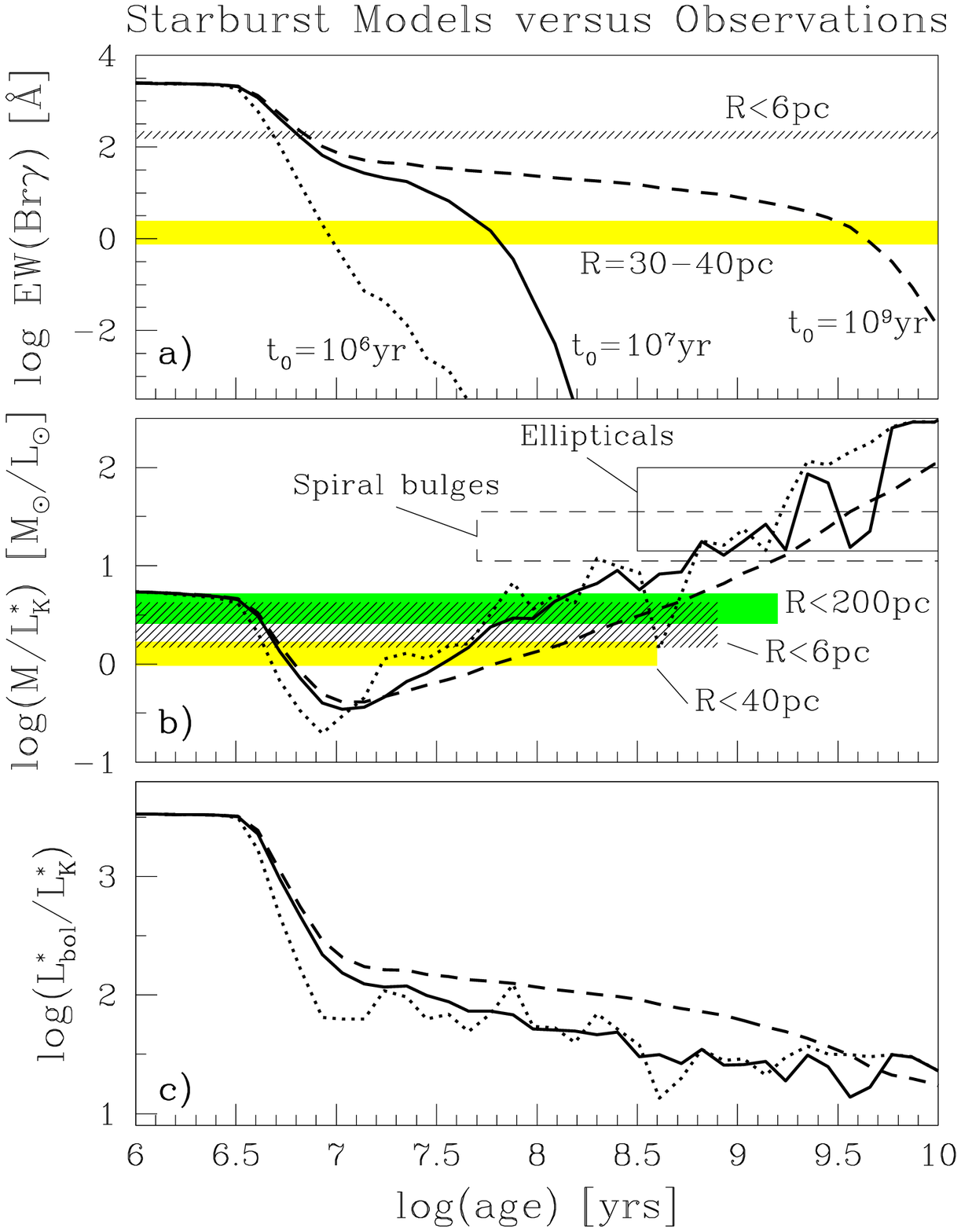}
\caption{}
\end{figure}

\begin{figure}
\figurenum{11}
\epsscale{1}
\plotone{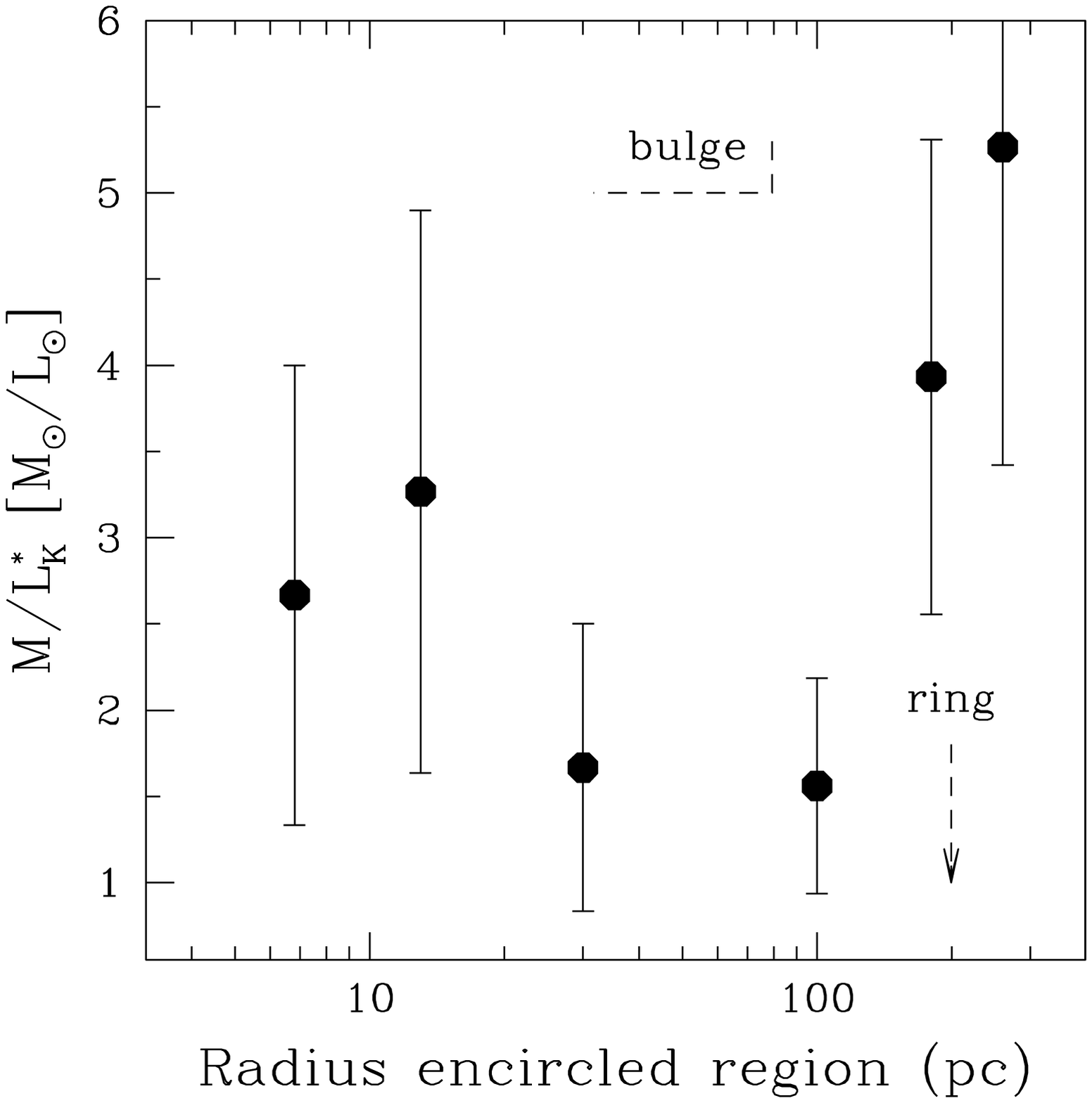}
\caption{}
\end{figure}

\begin{figure}
\figurenum{12}
\epsscale{1}
\plotone{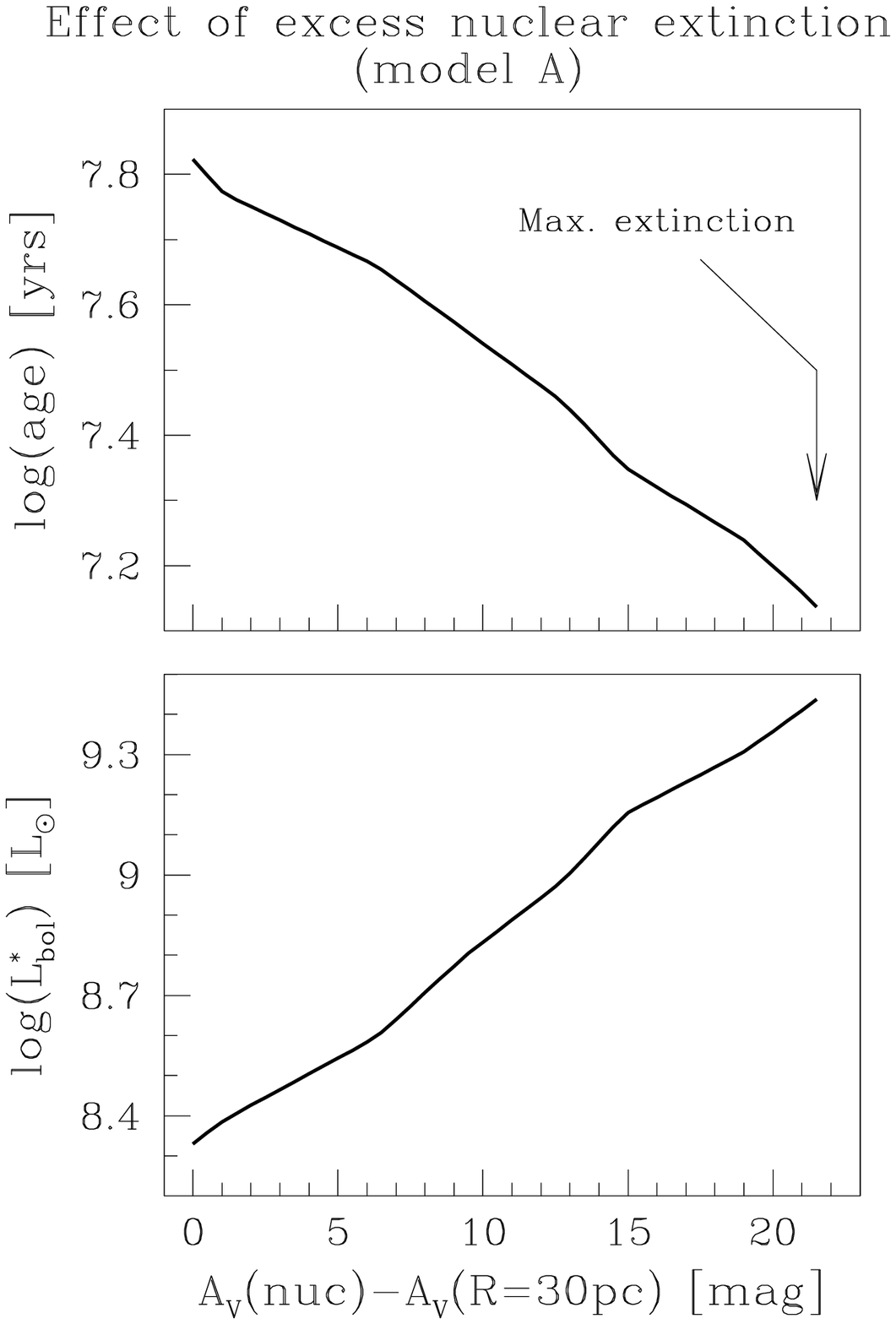}
\caption{}
\end{figure}

\begin{figure}
\figurenum{13}
\epsscale{1}
\plotone{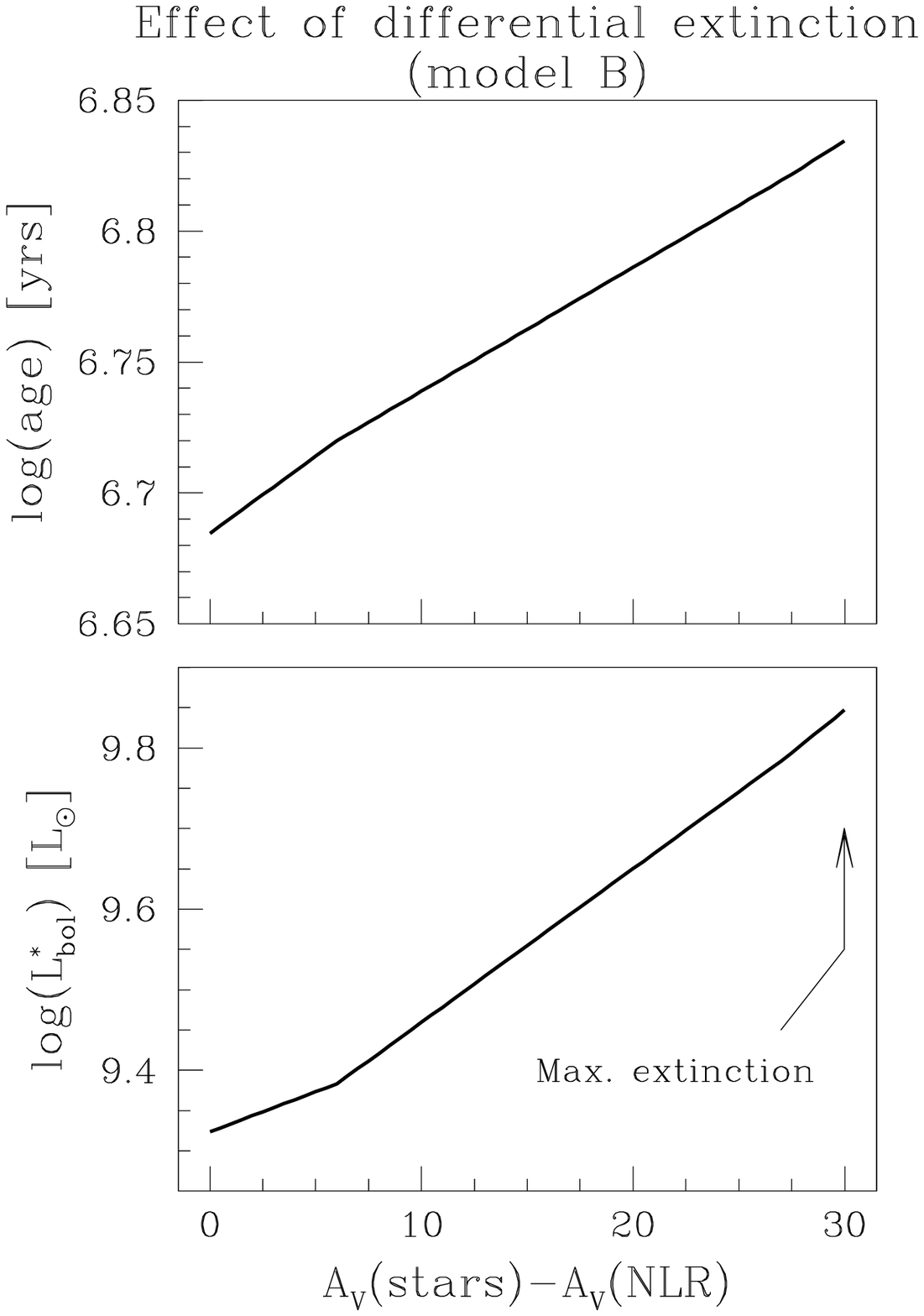}
\caption{}
\end{figure}

\begin{figure}
\figurenum{14}
\epsscale{1}
\plotone{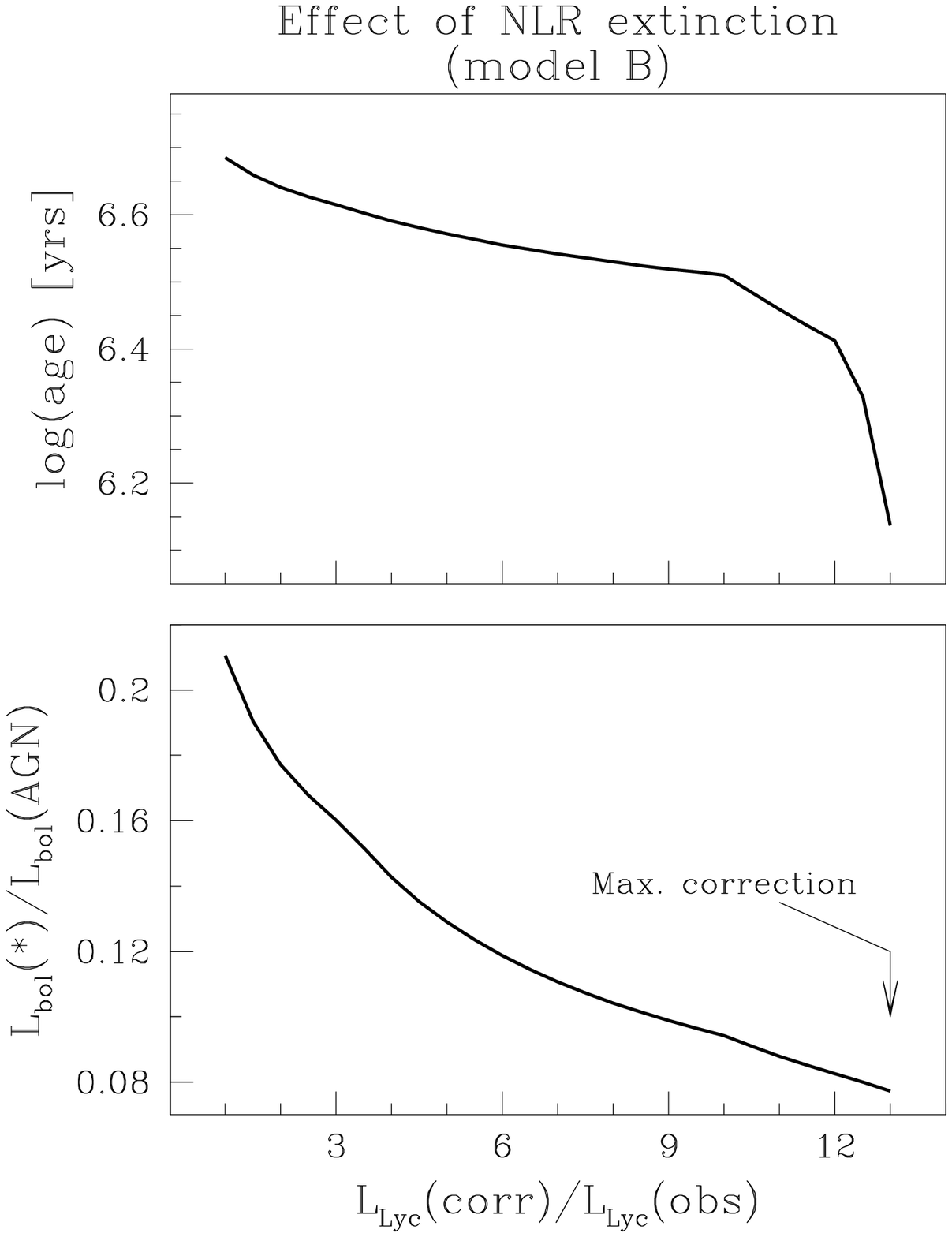}
\caption{}
\end{figure}

\end{document}